\documentclass[sigconf, nonacm]{acmart}
\usepackage{tikz}
\usepackage[linesnumbered,ruled,noend]{algorithm2e}
\usepackage{graphicx}
\usepackage{booktabs}
\usepackage{caption}
\usepackage{makecell}
\usepackage{fontawesome5}
\usepackage{amsmath,mathtools,booktabs,tabularx}

\usetikzlibrary{positioning,calc}
\usetikzlibrary{decorations.text}
\usetikzlibrary{decorations.pathmorphing,decorations.pathreplacing}
\usetikzlibrary{arrows,petri, topaths,fit}
\usetikzlibrary{shadows}

\usepackage{pgf}
\usepackage{pgfplots}
\usepgfplotslibrary{groupplots}
\pgfplotsset{compat=newest}

\usepackage{adjustbox}

\usepackage[utf8]{inputenc}
\usepackage{xcolor}        
\definecolor{mplblue}{HTML}{1f77b4}
\usepackage{listings}      

\lstdefinelanguage{Rust}{
  morekeywords={
    abstract,alignof,as,become,box,break,const,continue,crate,do,else,enum,
    extern,false,final,fn,for,if,impl,in,let,loop,macro,match,mod,move,mut,
    offsetof,override,priv,proc,pub,pure,ref,return,self,Self,sizeof,static,
    struct,super,trait,true,type,typeof,unsafe,unsized,use,virtual,where,while,yield, 
  },
  morekeywords=[2]{
    arrange, map, filter, flat_map, inspect, distinct, concat, join, join_map, join_core, semijoin, iterate
  },
  sensitive=true,
  morecomment=[l]{//},
  morecomment=[s]{/*}{*/},
  morestring=[b]{"}
}

\lstdefinelanguage{Datalog}{
  morekeywords={
    reach, h, two_hops, VarPointsTo, ArrayIndexPointsTo, LoadArrayIdx, SupertypeOf, Reach, p, q, p1, q1, p2, q2, p3, c4, CC, red, blue, answer
  },
  morekeywords=[2]{
    edge, node, target 
  },
  sensitive=true,
  morecomment=[l]{//},
  morecomment=[s]{/*}{*/},
  morestring=[b]{"}
}

\lstdefinestyle{DatalogStyle}{
  language=Datalog,
  keywordstyle=\color{black}\bfseries,    
  keywordstyle=[2]\color{black},            
}

\usepackage{wrapfig}
\usepackage{framed}   
 {\endMakeFramed}
\definecolor{shadecolor}{gray}{0.75}
\usepackage{thmtools, thm-restate}

\usepackage{multirow}
\usepackage{subcaption}
\usepackage{url}
\usepackage{graphicx}
\usepackage{tcolorbox}
\usepackage{enumitem}
\usepackage{array}
\usepackage{makecell}
\usepackage{pgfplots}
\pgfplotsset{compat=1.17}
\usepackage{xcolor}
\usepackage[table]{xcolor}
\definecolor{purple}{RGB}{128,0,128}
\definecolor{navy}{RGB}{0,0,128}
\definecolor{skyblue}{RGB}{135,206,235}
\definecolor{darkorange}{RGB}{255,140,0}
\definecolor{firebrick}{RGB}{178,34,34}

\newcommand{\thickcross}{\textcolor{red!60!black}{\tikz[baseline=-0.02em]{\draw[line width=1.5pt] (0,0) -- (0.12,0.12); \draw[line width=1.5pt] (0,0.12) -- (0.12,0);}}}

\newcommand*\circledwhite[1]{\tikz[baseline=(char.base)]{
            \node[shape=circle,fill=gray,text=white,inner sep=1pt] (char) {#1};}}

\newcommand{\high}[1]{{\textcolor{blue}{\textbf{\noindent{#1}}}}}
\newcommand{\higher}[1]{{\textcolor{red}{\textbf{\noindent{#1}}}}}

\definecolor{revonecolor}{RGB}{230,126,34}    
\definecolor{revtwocolor}{RGB}{39,174,96}     
\definecolor{revthreecolor}{RGB}{155,89,182}  
\definecolor{revallcolor}{RGB}{139,0,0}       

\newcommand{\introparagraph}[1]{\noindent \textbf{#1.}}
\newcommand{\Datalog}{\text{Datalog}\xspace}

\newcommand{\IDB}{\textsc{idb}\xspace}
\newcommand{\EDB}{\textsc{edb}\xspace}
\newcommand{\dbms}{\textsc{dbms}\xspace}

\newcommand{\eclair}{\textsf{FlowLog}\xspace}
\newcommand{\souffle}{\text{Soufflé}\xspace}
\newcommand{\ddlog}{\text{DDlog}\xspace}
\newcommand{\recstep}{\text{RecStep}\xspace}

\newcommand{\ir}{IR\xspace}

\newcommand{\IDBs}{\textsc{idb}s\xspace}
\newcommand{\EDBs}{\textsc{edb}s\xspace}

\newcommand{\sip}{\texttt{sip}\xspace}

\newcommand{\joinop}{\texttt{Join}\xspace}
\newcommand{\mapop}{\texttt{Map}\xspace}
\newcommand{\flatop}{\texttt{FlatMap}\xspace}

\newcommand{\filterop}{\texttt{Filter}\xspace}

\usepackage{hyperref}

\newcommand\vldbdoi{XX.XX/XXX.XX}
\newcommand\vldbpages{XXX-XXX}
\newcommand\vldbvolume{19}
\newcommand\vldbissue{3}
\newcommand\vldbyear{2025}
\newcommand\vldbauthors{\authors}
\newcommand\vldbtitle{\shorttitle} 
\newcommand\vldbavailabilityurl{https://github.com/hdz284/FlowLog}
\newcommand\vldbpagestyle{empty} 

\begin{document}
\title{FlowLog: Efficient and Extensible Datalog via Incrementality}

\author{Hangdong Zhao}
\affiliation{Microsoft Gray Systems Lab
}
\email{hangdongzhao@microsoft.com}

\author{Zhenghong Yu}
\affiliation{University of Wisconsin, Madison
}
\email{zyu379@wisc.edu}

\author{Srinag Rao}
\affiliation{University of Wisconsin, Madison
}
\email{srinskit@cs.wisc.edu}

\author{Simon Frisk}
\affiliation{University of Wisconsin, Madison
}
\email{simon.frisk@wisc.edu}

\author{Zhiwei Fan}
\affiliation{Meta Platforms Inc.
}
\email{zhiweifan@meta.com}

\author{Paraschos Koutris}
\affiliation{University of Wisconsin, Madison
}
\email{paris@cs.wisc.edu}

\begin{abstract}
Datalog-based languages are regaining popularity as a powerful abstraction for expressing recursive computations in domains such as program analysis and graph processing. However, existing systems often face a trade-off between efficiency and extensibility. Engines like \souffle achieve high efficiency through domain-specific designs, but lack general-purpose flexibility. Others, like \recstep, offer modularity by layering \Datalog on traditional databases, but struggle to integrate \Datalog-specific optimizations. 

This paper bridges this gap by presenting \eclair, a new Datalog engine that uses an explicit relational \ir per-rule to cleanly separate recursive control (e.g., semi-naïve execution) from each rule's logical plan. This boundary lets us retain fine-grained, \Datalog-aware optimizations at the logical layer, but also reuse off-the-shelf database primitives at execution. At the logical level (i.e. \ir), we apply proven SQL optimizations, such as logic fusion and subplan reuse. To address high volatility in recursive workloads, we adopt a robustness-first approach that pairs a structural optimizer (avoiding worst-case joins) with sideways information passing (early filtering). Built atop Differential Dataflow—a mature framework for streaming analytics—\eclair supports both batch and incremental \Datalog and adds novel recursion-aware optimizations called Boolean (or algebraic) specialization. Our evaluation shows that \eclair outperforms state-of-the-art Datalog engines and modern databases across a broad range of recursive workloads, achieving superior scalability while preserving a simple and extensible architecture.
\end{abstract}

\maketitle

\pagestyle{\vldbpagestyle}
\begingroup\small\noindent\raggedright\textbf{PVLDB Reference Format:}\\
\vldbauthors. \vldbtitle. PVLDB, \vldbvolume(\vldbissue): \vldbpages, \vldbyear.\\
\href{https://doi.org/\vldbdoi}{doi:\vldbdoi}
\endgroup
\begingroup
\renewcommand\thefootnote{}\footnote{\noindent
This work is licensed under the Creative Commons BY-NC-ND 4.0 International License. Visit \url{https://creativecommons.org/licenses/by-nc-nd/4.0/} to view a copy of this license. For any use beyond those covered by this license, obtain permission by emailing \href{mailto:info@vldb.org}{info@vldb.org}. Copyright is held by the owner/author(s). Publication rights licensed to the VLDB Endowment. \\
\raggedright Proceedings of the VLDB Endowment, Vol. \vldbvolume, No. \vldbissue\ %
ISSN 2150-8097. \\
\href{https://doi.org/\vldbdoi}{doi:\vldbdoi} \\
}\addtocounter{footnote}{-1}\endgroup

\ifdefempty{\vldbavailabilityurl}{}{
\vspace{.3cm}
\begingroup\small\noindent\raggedright\textbf{PVLDB Artifact Availability:}\\
The source code, data, and/or other artifacts have been made available at \url{\vldbavailabilityurl}.
\endgroup
}

\section{Introduction}

The rapid expansion of data-intensive applications has underscored the need for query languages that are both simple and expressive. As a declarative language, \Datalog adds recursion to relational algebra in a concise syntax, making it especially well-suited for domains such as graph processing~\cite{SociaLite13, recstep}, network monitoring~\cite{Abiteboul05}, program analysis~\cite{Bernhard2016, Szabo2016}, and distributed systems~\cite{Chu2024}. 



In recent years, academic advances and industry demands have spurred the development of \Datalog engines. In program analysis, \Datalog has proven effective for static analysis tasks such as bug detection and security checks, inspiring systems such as \souffle~\cite{Bernhard2016}, Flix~\cite{Flix16}, Flan~\cite{Flan24} and Ascent~\cite{Ascent}, often with domain-specific optimizations~\cite{IndexSouffle, TrieSouffle}. While successful in their target domains, these systems tend to sacrifice flexibility: adding incremental maintenance for continuous updates typically requires major system modifications~\cite{David2021}, and scaling up/out to accommodate larger datasets frequently demands extensive engineering.

Other systems avoid building from scratch by layering \Datalog on existing databases. A prominent example is \recstep~\cite{recstep}, which compiles \Datalog into SQL and executes it in QuickStep~\cite{QuickStep} one iteration at a time. inheriting its optimizer and parallelism. However, relying on a black-box \dbms complicates \Datalog-specific optimizations, particularly those spanning multiple iterations, such as semi-naïve evaluation and index maintenance. While \recstep interjects between iterations to impose some of these optimizations, the back-and-forth control flow incurs non-negligible synchronization overhead. Similarly, \ddlog~\cite{ddlog} compiles \Datalog directly into Differential Dataflow\footnote{Differential dataflow~\cite{McSherryMII13} programs chain up a set of DD's streaming operators that continuously maintain states for efficient incremental computation as data evolves (see Sec.~\ref{sec:background_dd} for a formal introduction.)} (DD) programs. However, empirical studies report substantial memory overhead--often orders of magnitude higher than alternatives~\cite{MansurWC23, David2021, souffleInterp, dyck22}. 

Balancing flexibility and efficiency for \Datalog remains an open challenge. This paper addresses this by pursuing a new design that unifies $(i)$ efficient, off-the-shelf relational operators as execution primitives, and $(ii)$ flexible, fine-grained optimization controls. To achieve the first goal, we reuse \textsf{DD}'s streaming operators as building blocks, positioning \ddlog as a baseline that directly code-generates \Datalog into lower-level \textsf{DD} without a distinct optimization phase. Achieving the second goal requires a novel design for \Datalog optimization. There are two main reasons for this: $(i)$ the primary bottleneck of \ddlog is memory usage, and $(ii)$ recursion weakens many conventional SQL techniques—e.g., cost-based planning~\cite{LeisGMBK015} lacks reliable static statistics in recursive contexts—so existing systems (e.g., \souffle, \ddlog) fall back on ad hoc heuristics or manual performance tuning. To curb memory usage, we apply a suite of memory-focused rewrites on a relational intermediate representation (\ir) that fully separates the rule's logical plan from its physical realization. Additionally, instead of the conventional cost-based planning, we opt for robustness-aware optimizations, which prioritizes avoiding worst-case scenarios (e.g., pathological join orders and large intermediate relations) over seeking a best query plan. 

These pieces culminate in \eclair: a high-performance \Datalog system targeting both batch and incremental processing. Its simple architecture eases integration of novel optimizations (e.g. Boolean specialization, Sec.~\ref{sec:specialization}), rich semantics (e.g. recursive aggregations) and scale-out extensions (Sec.~\ref{sec:extensibility}) with minimal system changes. In summary, this paper makes the following \underbar{\textbf{contributions}}:

\begin{enumerate}
  \item \textbf{System Architecture.} Sec.~\ref{sec:design} presents the new system design of \eclair, centered around an IR that decouples ever rule's logical structure from its physical execution in \textsf{DD}. While this logical/physical split is long standard in modern SQL databases, it has been largely absent from existing \Datalog engines, limiting systematic \Datalog optimization.
  \item \textbf{Optimization \& Robustness.} \eclair integrates a suite of \ir-level (i.e. logical) optimizations, including proven SQL techniques such as logic fusion (Sec.~\ref{sec:fusion}) and subplan sharing (Sec.~\ref{sec:sharing}) to shrink \textsf{DD}'s state and memory footprint. Sec.~\ref{sec:planning} presents \eclair's optimizer that analyzes the join graph of each rule to avoid pathological join orders. Sec.~\ref{sec:robust} complements this by semijoin pre-filtering that stabilizes the execution of recursive workloads. Both techniques (Sec.~\ref{sec:planning}-\ref{sec:robust}) are geared towards mitigating the inherent high volatility and unpredictability we observed in \Datalog workloads.
  \item \textbf{Extensions.} Sec.~\ref{sec:specialization}-\ref{sec:extensibility} outline how \eclair's modular design supports $(i)$ incremental maintenance, $(ii)$ novel \Datalog-aware optimizations (i.e., Boolean/algebraic specialization for recursive aggregation), and $(iii)$ scale-out execution.
  \item \textbf{Experiments.} Sec.~\ref{sec:experiments} conducts extensive experiments on a broad set of benchmarks we have collected across recent literature and open-source projects, spanning multiple domains and heterogeneous workload characteristics. Our results show that \eclair often substantially outperforms state-of-the-art Datalog engines and modern databases, in terms of latency, memory usage, and scalability.
\end{enumerate}

\introparagraph{Related Work}
As~\cite{recstep, DBLP:journals/ftdb/KetsmanK22} pointed out, most \Datalog engines are purpose-built for specific domains~\cite{Bernhard2016, souffleInterp, Szabo2016, Flix16, Ascent, Flan24, DBLP:conf/icde/HerlihyMAO24, DBLP:conf/sigmod/ShkapskyYICCZ16, GPUDatalog, shovon2025multi}. Over time, they have introduced a range of optimizations and extensions tailored to \Datalog~\cite{souffleInterp, LogicBlox, NestedRec24, DCDatalog22,datalogo, egglog, 10.14778/3712221.3712232, DBLP:conf/datalog/ZhaoKD24}, such as incremental \Datalog~\cite{David2021, ddlog}. We incorporate some of these techniques into \eclair (e.g., index sharing~\cite{IndexSouffle, arrangement}, unified \IDB evaluation~\cite{recstep}), while others, such as magic sets~\cite{WangK0PS22}, de-specialized relations~\cite{souffleInterp}, customized data structures~\cite{DBLP:journals/pacmpl/SahebolamriBMM23} and worst-case optimal joins~\cite{Flan24}, remain promising future explorations. Notably, major gaps in the \Datalog literature persist, particularly the lack of effective query planning and limited scalability in highly iterative workloads, both being critical challenges for large-scale \Datalog applications~\cite{DBLP:conf/lopstr/ArchHZSS22, DBLP:conf/datalog/FanMK22}.

\section{Background}
In this section, we provide a brief overview of standard \Datalog, its evaluation, and common extensions.

\subsection{\Datalog Basics}
A standard \Datalog program~\cite{abiteboul1995foundations} is a set of {\em rules}. A rule is an expression of the following form:

\begin{lstlisting}[mathescape=true, numbers=none, basicstyle=\footnotesize\ttfamily]
    h :- p1, p2, ..., pk.
\end{lstlisting}
The terms \texttt{\footnotesize h, p1, \ldots, pk} are atoms, i.e., formulas of $R(x, y, \ldots)$, where $R$ is the atom (or relation) name and $(x, y, \ldots)$ is its variables (or attributes). The atom $h$ is the {\em head} and the atoms $p_1, \ldots, p_k$ are the {\em body} of the rule. A rule can be interpreted as a logical implication: if $p_1, \ldots, p_k$ are true, then so is the head $h$. We assume that every attribute of $h$ occurs in some $p_i$ The atoms of a \Datalog program are of two types: $\IDB$ and $\EDB$. An atom that represents an input relation is an \EDB (extensional database); an \EDB comprises a set of (base) facts/tuples and is never the head of a rule. A atom that represents a derived relation is an \IDB (intensional database, in \textbf{bold} font); an \IDB must appear in the head of at least one rule.

\begin{example} \label{ex:reach}
Consider the following task over a directed graph: find all nodes that can reach a target via an even number of hops. We represent the graph as a binary \EDB $\textsf{edge}(x, y)$, where $(x, y)$ is a fact if there is an edge from node $x$ to node $y$. We use another \EDB $\textsf{target}(x)$ as the unary relation containing the target node $\textsf{a}$. The task can then be expressed in \Datalog as follows:


\begin{lstlisting}[style=DatalogStyle, mathescape=true, numbers=none, basicstyle=\footnotesize\ttfamily]
    r1.  reach(x) :- target(x). 
    r2.  reach(x) :- edge(x,y), edge(y,z), reach(z).
\end{lstlisting}
Here, the atom $\textsf{reach}(x)$ is the \IDB to output. $r_1$ initializes the trivial case: the target node is reachable in zero hop. The second rule $r_2$ is recursive and states that if there is a length-2 path from $x$ to $z$, and $z$ can reach the target using an even number of hops, then so can $x$.
\end{example}


\introparagraph{Dependency Graph and Stratification}
A {\em dependency graph} of a \Datalog program is a directed graph where every rule is a node; there is an edge from rule $r_1$ to $r_2$ if the head of $r_1$ appears in the body of $r_2$. A rule is {\em recursive} if it belongs to a directed cycle, and {\em non-recursive} otherwise. A {\em stratification} of a program is a partition of the rules into strata, where each stratum is the set of rules that belongs to the same strongly connected component of the dependency graph. The topological order of the strongly connected components defines the order among the strata. The dependency graph for Example~\ref{ex:reach} has two nodes $r_1$ and $r_2$, and edges $r_1 \rightarrow r_2$ and $r_2 \rightarrow r_2$. Thus, the program has strata, $\{r_1\}$ and $\{r_2\}$ in the topological order.

\smallskip
\introparagraph{Common \Datalog Extensions}
To enrich \Datalog for practical usage, we incorporate common syntactic extensions as~\cite{recstep}--constraints, stratified negations (where negated atoms are either \EDB or an \IDB from a lower stratum), and (possibly recursive) aggregations. For example, this allows finding two hops $(x, z)$ in a graph that are:
\begin{lstlisting}[mathescape=true, numbers=none, basicstyle=\footnotesize\ttfamily]
    edge(x,y), edge(y,z), x $\neq$ z       // not loops
    edge(x,y), edge(y,z), $\neg$edge(x,z)  // not one hop
\end{lstlisting}
or for each $x$ and $z$, counting the number of possible two hops:
\begin{lstlisting}[style=DatalogStyle, mathescape=true, numbers=none, basicstyle=\footnotesize\ttfamily]
    two_hops(x,z,COUNT(y)) :- edge(x,y), edge(y,z).
\end{lstlisting}


\subsection{\Datalog Evaluation}
The straightforward way to implement \Datalog is via {\em naïve} evaluation. Starting with the set of all \EDB facts, we iteratively apply every rule as a join query to derive new facts, adding them to the head \IDBs until no new facts can be derived, i.e., a fixpoint is reached.

However, naïve evaluation is usually wasteful because each iteration executes rules on all historical data, leading to rediscovery of facts derived in previous iterations. Hence, modern \Datalog engines use {\em semi-naïve} evaluation, which only uses new tuples from the last iteration to derive facts in the current iteration. A common practice is to further exploit stratification: each stratum gets evaluated in order, and the results are used as input for the next stratum. In Example~\ref{ex:reach}, the first stratum simply inserts the target node $t$ to $\textsf{reach}(x)$. Next, at each iteration $i = 1, 2, \ldots$, we only consider the new facts derived in the $(i-1)$-th iteration, denoted as $\Delta \textsf{reach}^{i}$ where $\Delta \textsf{reach}^{0} = \{\textsf{a}\}$, to populate new facts by computing the join $\textsf{edge}(x, y) \bowtie \textsf{edge}(y, z) \bowtie \Delta \textsf{reach}^{(i-1)}$.


\subsection{Differential Dataflow} \label{sec:background_dd}
Differential Dataflow (or \textsf{DD})~\cite{McSherryMII13} is a data-parallel programming model for large-scale incremental data processing. Its Rust implementation\footnote{\url{https://github.com/TimelyDataflow/differential-dataflow}} compiles down to Timely Dataflow, a lower-level generic distributed streaming system introduced by ~\cite{Timely}. 

\smallskip
\introparagraph{Collections} 
\textsf{DD} abstracts a relation as a stream of rows, termed as a collection. A row in a collection is a triple $(\texttt{data}, \texttt{time}, \texttt{diff})$, where \texttt{data} is the raw tuple from the relation, \texttt{time} is the timestamp when the tuple is ingested, and \texttt{diff} is its multiplicity. The \texttt{diff} field is for \textsf{DD} to annotate duplications and track incremental changes (i.e., $+\delta$ represents an insertion of $\delta$ copies of the tuple and $-\delta$ represents a retraction of $\delta$ copies). In Example~\ref{ex:reach}, \textsf{DD} constructs corresponding input collections for $\textsf{target}(x)$ and $\textsf{edge}(x, y)$, where the former is initialized as a single row $(\textsf{a}, 0, 1)$ for the target $\textsf{a}$; the latter is a collection of rows $((\textsf{a}, \textsf{b}), 0, 1)$ for each edge $(\textsf{a}, \textsf{b})$ of the graph. Here, $0$ is the initial timestamp. 

\smallskip
\introparagraph{Differential Operators} 
 \textsf{DD} uses a set of incremental operators such as \textcolor{blue}{\texttt{map}}, \textcolor{blue}{\texttt{filter}}, and \textcolor{blue}{\texttt{join}}, that each imposes a relational operation on input collection(s) and outputs a collection. We can compose them to compute \texttt{SQL} queries and quickly respond to input changes. At its core, every differential operator maintains a minimal set of changes to the output when the input changes, and propagates these updates in a semi-naïve manner -- only considering changes since the last timestamp. Let $R(x, y)$ and $S(x, z)$ be two collections. Suppose $((\textsf{a}, \textsf{b}), 0, 4)$ and $((\textsf{c}, \textsf{b}), 0, 1)$ are two rows of $R(x, y)$ and $((\textsf{a}, \textsf{d}), 0, 3)$ is a row of $S(x, z)$. A rule \texttt{T(y,z) :- R(x,y), S(x,z)} maps to a composition of \textcolor{blue}{\texttt{join}} and \textcolor{blue}{\texttt{map}} as follows:
 \begin{lstlisting}[mathescape=true, numbers=none, basicstyle=\footnotesize\ttfamily]
    R.join(&S).map(|t| (t.y, t.z));
\end{lstlisting}
 Here, \textcolor{blue}{\texttt{join}} emits a row $((\textsf{a}, (\textsf{b}, \textsf{d})), 0, 12)$, where the output \texttt{data} is formatted as $(x, (y, z))$ (i.e. join keys followed by a grouping of values) and \texttt{diff} is the product of two input \texttt{diff}s, i.e., $4 \times 3 = 12$. The downstream \textcolor{blue}{\texttt{map}} projects to $(y, z)$ and gets $((\textsf{b}, \textsf{d}), 0, 12)$. If there is an insertion of $((\textsf{a}, \textsf{b}), 0, +2)$ in the collection, only a delta change $((\textsf{b}, \textsf{d}), 0, +2 \times 3 = +6)$ will propagate through.
 %

\textsf{DD} allows users to define custom operators (beyond standard \texttt{SQL}) without worrying about low-level incremental mechanisms. This makes \textsf{DD} a powerful backend for \Datalog applications. A unique but essential operator is \textcolor{blue}{\texttt{iterate}}, which repeatedly applies an enclosed \textsf{DD} closure to input collections. The following code snippet shows how to implement Example~\ref{ex:reach} as a \textsf{DD} program.

\begin{lstlisting}[numbers=none, basicstyle=\footnotesize\ttfamily]
    // (1) iterate starting from target
    target.iterate(|reach| {
        // (2) derive new reach(x) from joins
        edge.map(|t| (t.y, t.x))
            .join(&edge)
            .map(|t| (t.z, t.x))
            .join(&reach)
            .map(|t| t.x)
            // (3) concat and dedup for convengence
            .concat(&reach) 
            .distinct()           });
\end{lstlisting}

The \textcolor{blue}{\texttt{iterate}} operator initializes $\textsf{reach}$ as $(\textsf{a}, (0, 0), 1)$, i.e. \textsf{target}, and sets a series of nested timestamps $(0, i)$, where $0$ is the outer timestamp and $i = 0, 1, \ldots$ is the iteration counter. Then, it repeatedly applies the inner \textsf{DD} logic as $i$ increments--for each $i$, the \textcolor{blue}{\texttt{concat}} operator adds new output into $\textsf{reach}$ and \textcolor{blue}{\texttt{distinct}} de-duplicates the results, i.e. for each row, \texttt{diff} maps to $1$ if $\texttt{diff} > 1$. When no new rows are derived (i.e. fixpoint), \textcolor{blue}{\texttt{iterate}} collects all results and returns the final $\textsf{reach}$ to the outer scope/timestamp $0$.

The inner logic is verbose but necessary. An idiosyncratic feature of \textsf{DD} is the use of \textcolor{blue}{\texttt{map}} and \textcolor{blue}{\texttt{join}}; \textcolor{blue}{\texttt{map}} is not only a projection, but also a way to re-organize \texttt{data} into a key-value pair, e.g., the first \textcolor{blue}{\texttt{map}} swaps $(x, y)$ to $(y, x)$, and that designates $y$ as the key and $x$ as the value. This is because the \textcolor{blue}{\texttt{join}} operator of \textsf{DD} requires its two input collections to be explicitly pre-indexed on the join keys, using an \textcolor{blue}{\texttt{arrange}} operator described next. 

\smallskip
\introparagraph{Arrangements} 
An arrangement is an in-memory index for \textsf{DD} collections~\cite{arrangement}. It can be considered as a sorted dictionary that allows efficient concurrent access. An arrangement indexes batches of historical changes, maintains them over time, and merges them into compact representations as appropriate (e.g., when a timestamp is advanced). The first \textcolor{blue}{\texttt{join}} in the above code pre-arranges both operands by imposing an \textcolor{blue}{\texttt{arrange}} operation internally for each and then uses a more primitive \textcolor{blue}{\texttt{join\_core}} operator to join on arrangements. Indeed, the first \textcolor{blue}{\texttt{join}} is executed under the hood as
\begin{lstlisting}[numbers=none, basicstyle=\footnotesize\ttfamily]
    edge.map(|t| (t.y, t.x))
        .arrange()          // k: (y), v1: (x)
        .join_core(
            &edge.arrange() // k: (y), v2: (z)
        ); // output schema k: (y), v3: (x, z)
\end{lstlisting}

\section{System Design}
\label{sec:design}

This section presents the architecture of \eclair. which builds on \textsf{DD} as its execution backend. Instead of compiling \Datalog directly into a \textsf{DD} program, \eclair first translates rules into an intermediate representation (\ir) that  captures the logical structure of the program, then lowers to \textsf{DD}. This allows us to reason about optimizations purely at the logical level while abstracting away low-level details such as recursive control, de-duplication, and incrementality, making it simpler to modularize and extend. The \ir is akin to recent {\em grounding} approaches for \Datalog~\cite{datalogo, grounding24} and we incorporate these ideas when optimizing the \ir (see Section~\ref{sec:planning}).

\begin{figure}[t]
  \centering
  \includegraphics[width=0.8\linewidth]{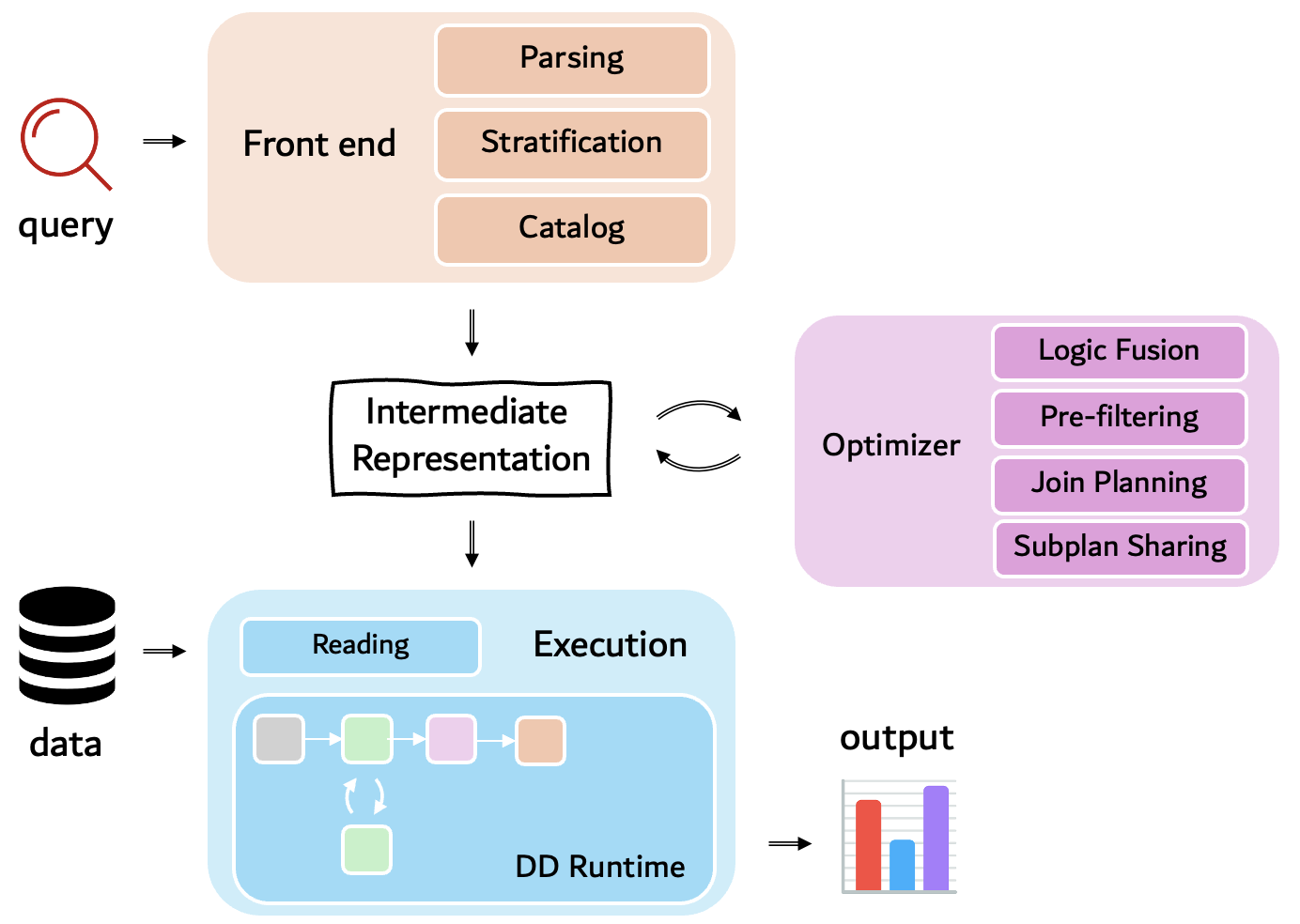}
  \caption{\small System Architecture of \eclair}
  \label{fig:eclair}
\end{figure} 

\smallskip
\introparagraph{Overview}
The architecture of \eclair has three components: the front-end, the optimizer, and the executor. The \textbf{front-end} parses the input \texttt{.dl} program (using \souffle grammar), performs syntax checking, stratifies the rules, and creates a per-rule catalog with  meta information such as the join graphs (formally defined in Section~\ref{sec:search_space}). The \textbf{optimizer} populates an \ir from the catalog and applies optimizations. The \textbf{executor} renders the optimized \ir to a \textsf{DD} program, reads the input data, and runs the iterative execution. 

\smallskip
\introparagraph{Query Optimization (\ir)}
For every rule in the input \Datalog program, \eclair's optimizer constructs an \ir. The \ir is a tree of logical transformations for that rule (i.e. a relational logical plan), e.g. \joinop, \mapop, and \filterop. Leaf nodes are input tables and intermediate nodes are logical transformations. Edges denote the data flowing across these transformations, and edge labels indicate the underlying schema. We insist every \joinop's inputs to align on its join keys so that \textsf{DD}'s physical join operator can consume it directly. 

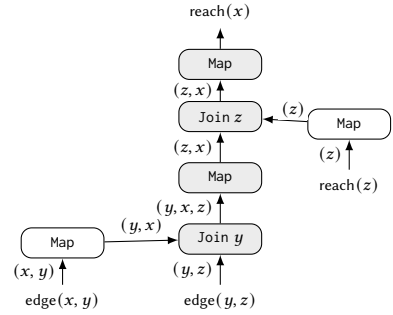
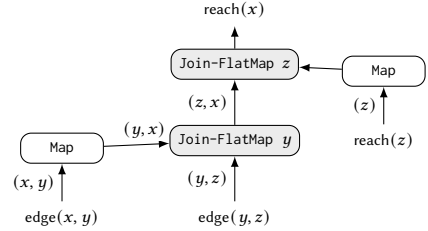
\begin{figure}[h]
  \centering
  
  \begin{subfigure}[t]{0.45\textwidth}
    \centering
    \begin{tikzpicture}[
        >=latex, 
        auto,
        node distance=0.4cm and 0.6cm,
        op/.style={
          draw,
          rounded corners,
          minimum width=11mm,
          minimum height=4mm,
          font=\scriptsize
        },
        fusedop/.style={
          draw,
          rounded corners,
          minimum width=11mm,
          minimum height=4mm,
          font=\scriptsize,
          fill=lightgray!30
        },
        labelstyle/.style={
          font=\scriptsize,
          midway,
          fill=white,
          inner sep=1pt
        },
        dottedbox/.style={
          draw,
          dotted,
          inner sep=4pt,
          rounded corners
        }
    ]

    \node[font=\scriptsize] (tbl_xy) at (0.5, 1.3) {\(\textsf{edge}(x,y)\)};
    \node[font=\scriptsize, right=0.9cm of tbl_xy] (tbl_yz) {\(\textsf{edge}(y,z)\)};
    
    \node[op, above=0.35cm of tbl_xy] (mapyz) {\mapop};
    \draw[->] (tbl_xy) -- node[labelstyle, xshift=-0.05cm]{$(x,y)$} (mapyz);

    \node[fusedop, above=0.4cm of tbl_yz] (joiny) {\joinop \ $y$};
    \draw[->] (mapyz) -- node[labelstyle, xshift=0cm, yshift=0.05cm]{$(y, x)$} (joiny);
    \draw[->] (tbl_yz) -- node[labelstyle, xshift=-0.05cm]{$(y, z)$} (joiny);
    
    \node[fusedop, above=0.4cm of joiny] (projxz) {\mapop};
    \draw[->] (joiny) -- node[labelstyle, xshift=-0.05cm]{$(y, x, z)$} (projxz);

    \node[fusedop, above=0.4cm of projxz] (joinz) {\joinop \ $z$};
    \draw[->] (projxz) -- node[labelstyle, xshift=-0.05cm]{$(z, x)$} (joinz);

    \node[op, right=0.6cm of joinz, yshift=-0.1cm] (mapz){\mapop};
    \draw[<-] (joinz) -- node[labelstyle, xshift=-0.09cm]{$(z)$} (mapz);

    \node[font=\scriptsize, below=0.4cm of mapz] (tbl_rz) {\(\textsf{reach}(z)\)};
    \draw[->] (tbl_rz) -- node[labelstyle, xshift=-0.05cm]{$(z)$} (mapz);

    \node[fusedop, above=0.28cm of joinz] (projx) {\mapop};
    \draw[->] (joinz) -- node[labelstyle, xshift=-0.05cm]{$(z, x)$} (projx);

    \node[font=\scriptsize, above=0.28cm of projx] (final_reach) {\(\textsf{reach}(x)\)};
    \draw[->] (projx) -- (final_reach);


    \end{tikzpicture}
    \caption{\small Initial \ir for $r_2$, shaded parts are consecutive \joinop and \mapop.}
    \label{fig:ir}
  \end{subfigure}
  \hfill

  \begin{subfigure}[t]{0.45\textwidth}
    \centering
    \begin{tikzpicture}[
        >=latex, 
        auto,
        node distance=0.6cm and 0.8cm,
        op/.style={
          draw,
          rounded corners,
          minimum width=11mm,
          minimum height=4mm,
          font=\scriptsize
        },
        fusedop/.style={
          draw,
          rounded corners,
          minimum width=11mm,
          minimum height=4mm,
          font=\scriptsize,
          fill=lightgray!30
        },
        labelstyle/.style={
          font=\scriptsize,
          midway,
          fill=white,
          inner sep=1pt
        },
        dottedbox/.style={
          draw,
          dotted,
          inner sep=4pt,
          rounded corners
        }
    ]

    \node[font=\scriptsize] (tbl_xy) at (0, 1.5) {$\textsf{edge}(x,y)$};
    \node[font=\scriptsize, right=1.1cm of tbl_xy] (tbl_yz) {$\textsf{edge}(y,z)$};

    \node[op, above=0.5cm of tbl_xy] (mapyx) {\mapop};
    \draw[->] (tbl_xy) -- node[labelstyle, xshift=-0.05cm]{$(x, y)$} (mapyx);

    \node[fusedop, above=0.6cm of tbl_yz] (join_map_y) {\texttt{Join-FlatMap} \ $y$};

    \node[font=\scriptsize, right=0.6cm of join_map_y] (tbl_rz) {$\textsf{reach}(z)$};
    
    \node[op, above=0.5cm of tbl_rz] (mapreach) {\mapop}; 
    \draw[->] (tbl_rz) -- node[labelstyle, xshift=-0.05cm]{$(z)$} (mapreach);

    \draw[->] (mapyx) -- node[labelstyle, xshift=0.1cm, yshift=0.05cm]{$(y, x)$} (join_map_y);
    \draw[->] (tbl_yz) -- node[labelstyle, xshift=-0.05cm]{$(y,z)$} (join_map_y);

    \node[fusedop, above=0.6cm of join_map_y] (join_map_z) {\texttt{Join-FlatMap} \ $z$};

    \draw[->] (join_map_y) -- node[labelstyle, xshift=-0.05cm]{$(z, x)$} (join_map_z);
    \draw[->] (mapreach) -- (join_map_z);
    
    \node[font=\scriptsize, above=0.3cm of join_map_z] (final_reach) {$\textsf{reach}(x)$};
    \draw[->] (join_map_z) -- node[labelstyle, xshift=-0.05cm]{} (final_reach); 


    \end{tikzpicture}
    \caption{\small An optimized \ir by fusing \joinop and \mapop into \texttt{Join-FlatMap}.}
    \label{fig:fused_ir}
  \end{subfigure}
  \caption{\small Logic Fusion for $r_2$ from Example~\ref{ex:reach}}
  \label{fig:merged_ir}
\end{figure}

Figure~\ref{fig:ir} presents an \ir for $r_2$ (Example~\ref{ex:reach}). The lowest \mapop of $\textsf{edge}(x,y)$ assigns $y$ as the join key ($x$ as the value) for its parent \joinop, which binds both inputs to $y$ and pushes the join result of schema $(y, x, z)$ upstream; edge labels are omitted when the schema is obvious. The \ir always reads like an ordinary SQL query plan, and thus easily accepts a suite of relational optimizations.

\smallskip
\introparagraph{Query Execution}
The executor takes a set of (optimized) \ir (one for each input rule) and coalesces it into a global dataflow graph of (physical) differential operators, which are executed iteratively, strata by strata, until fixpoint. This execution taps into \textsf{DD}'s incremental and asynchronous nature, achieving efficiency and scalability out of the box. However, the incrementality asks every operator to maintain its state in memory. As exhibited by \ddlog against systems such as \souffle, the key overhead stems from these internal footprints, which can be prohibitive for large intermediate results. As such, Sec.~\ref{sec:fusion}-\ref{sec:sharing} will each present an \ir-level optimization (e.g. fusing transformations, re-using subplans) that are all geared towards minimizing intermediate output sizes, thus making \eclair much more competitive than \ddlog for large-scale recursive queries.


\section{Logic Fusion}
\label{sec:fusion}

Logic fusion is a query optimization technique that merges multiple small and adjacent logic (of frequent occurrence) into a single one to reduce dispatches during interpretation. For \eclair, logic fusion in addition eliminates unnecessary intermediate operator states. We describe two most effective fusion patterns for \eclair's \ir and assume that they will always be applied in the rest of the paper.

\smallskip
\introparagraph{Consecutive \mapop and \filterop} We introduce a new \flatop transformation into the \ir that fuses consecutive \mapop and \filterop (similar fusions are applied to optimize incremental \texttt{SQL}\footnote{https://materialize.com/blog/generalizing-linear-operators/\#fusing-logic}). A \flatop mirrors a lower-level \textcolor{blue}{\texttt{flat\_map}} physical operator of \textsf{DD} that filters and projects tuples to the desired schema in one pass. For example, the following rule and code snippet find the neighbors of node \textsf{a}:
\begin{lstlisting}[numbers=none, basicstyle=\footnotesize\ttfamily]
  // neighbor(y) :- edge(x, y), x = a.
  let neighbor = edge.flat_map(|t| 
    if t.x == a { vec![(t.y)] } else { vec![] } ); 
\end{lstlisting}


\smallskip
\introparagraph{\joinop followed by \mapop/\filterop} A \texttt{Join-FlatMap} fuses a \joinop with subsequent \mapop(s) and \filterop(s). It avoids materializing the full join output that is immediately projected or filtered. We will apply it as a default optimization for \eclair's \ir. The initial \ir of $r_2$ (Figure~\ref{fig:ir}) is optimized by fusing every consecutive \joinop and \mapop. In the fused \ir (Figure~\ref{fig:fused_ir}), the lower \texttt{Join-FlatMap} directly emits $(z, x)$ tuples; and the upper one now emits $(x)$ tuples instead of $(z, x)$. At the executor, \texttt{Join-FlatMap} renders into a \textcolor{blue}{\texttt{join\_core}} physical \textsf{DD} operator, which works as the following psuedocode:
\begin{lstlisting}[numbers=none, basicstyle=\footnotesize\ttfamily]
    edge.join_core(&edge, |t| 
        if some filters are passed // if any
            vec![(t.z, t.x)] // map to (z, x)
        else vec![]  );      // filtered out
\end{lstlisting}



\section{Join-project Plan Optimization} 
\label{sec:planning}
Traditional \dbms optimizers rely on approximate statistics to choose plans at the lowest cost~\cite{DBLP:journals/pvldb/LeisGMBK015}. However, finding optimal query plans for \Datalog rules is much more strenuous as these statistics are often missing or unstable: \IDB(s) accumulates at runtime in varying delta sizes across iterations, and real-world applications such as \textsf{DOOP}~\cite{BravenboerS09} and \textsf{DDISASM}~\cite{DDISASM} often involve highly complex join topologies (e.g. cyclic joins) than those typically handled by traditional \dbms. Hence systems \souffle and \ddlog give up and use only hard-coded listing orders: if a rule is written with body $R, S, T$, the system executes joins in that order even if there are cross products, i.e. $(R \bowtie S) \bowtie T$. \recstep collects runtime statistics periodically and invokes the \dbms's optimizer to re-optimize query plans on the fly, paying a synchronization and catalog maintenance overhead.

\eclair's optimizer chooses a static join–project plan per rule and maps it one-to-one to the \ir. Rather than relying on any runtime statistics, it analyzes the rule’s join graph using conventional heuristics (e.g., filter pushdown) plus worst-case-aware analysis~\cite{grounding24}. Although not always optimal, this approach reliably avoids disastrous orders in our experiments (Sec.~\ref{sec:experiments}) and provides a principled base for future \EDB-aware cost models.


\subsection{Structural Cost Model} \label{sec:model}
\eclair adopts a structural cost model that uses the distinct number of participating variables as a proxy for the asymptotic cost of a transformation. For example, in Figure~\ref{fig:fused_ir}, the lower \texttt{Join-FlatMap} involves $x, y, z$ variables and hence bears a costing of $3$, while the \texttt{Join-FlatMap} above involves $z, x$ and is assigned a cost of 2.

We define the cost of a join-project plan (or equivalently, an \ir) as the maximum cost of any transformation it contains. In Figure~\ref{fig:fused_ir}, the maximum is 3, determined by the lower \texttt{Join-FlatMap}. Intuitively, for Example~\ref{ex:reach} over an input graph of $n$ nodes, this plan implies a worst-case intermediate size (and so is the asymptotic runtime) of $O(n^{3})$. The same reasoning extends to arbitrary \Datalog programs: the structural cost upper-bounds worst-case intermediate output sizes, and formal guarantees are in~\cite{grounding24}. Similar techniques are proposed for subgraph pattern matching~\cite{joinwidth} and have shown effectiveness for multi-way many-to-many joins.

\subsection{Search Strategy} \label{sec:search_space}
We establish necessary definitions to describe the search space of join-project plans. The optimizer picks the plan in the search space that minimizes the cost under the structural cost model in Sec.~\ref{sec:model}.

A \textbf{(weighted) join graph} of a rule is a graph where every node is an atom and an edge exists if the two atoms join on at least one variable, and its weight is the number of variables they join on.


A \textbf{join tree}~\cite{Yannakakis81} is a spanning tree of the join graph such that for every variable $x$, the atoms containing $x$ (in its schema) induce a connected subtree of the spanning tree. The join tree is then used to define acyclicity of a rule, i.e. a rule is acyclic if and only if there is a join tree for it. Zhao et al.~\cite{grounding24} showed that for acyclic rules, bottom-up join orders (rooted arbitrarily) along a join tree, after early projections, usually yield tight intermediate size bounds. 

\begin{figure}[t]
  \centering
  \begin{tikzpicture}[
      >=latex, 
      auto,
      node distance=0.8cm and 1.6cm,
      op/.style={
        draw,
        rounded corners,
        minimum width=10mm,
        minimum height=3.5mm,
        font=\scriptsize
      },
      labelstyle/.style={
        font=\scriptsize,
        midway,
        fill=white,
        inner sep=0.8pt
      },
      dottedbox/.style={
        draw,
        dotted,
        inner sep=3pt,
        rounded corners
      }
  ]

  \node[font=\scriptsize] (jst_r4) at (0,1.2) {\(\textsf{edge}(x,y)\)};
  \node[font=\scriptsize, below=0.6cm of jst_r4] (jst_r5) {\(\textsf{edge}(y,z)\)};
  \node[font=\scriptsize, below=0.6cm of jst_r5] (jst_r6) {\(\textsf{reach}(z)\)};

  \draw[->] (jst_r5) -- node[labelstyle, xshift=0.4cm]{\circledwhite{2}} (jst_r4);
  \draw[->] (jst_r6) -- node[labelstyle, xshift=0.4cm]{\circledwhite{1}} (jst_r5);

  \node[font=\scriptsize, right=3.5cm of jst_r6] (tbl_rz) {\(\textsf{reach}(z)\)};
  \node[font=\scriptsize, left=1.2cm of tbl_rz] (tbl_yz) {\(\textsf{edge}(y,z)\)};
  \node[font=\scriptsize, above=1.2cm of tbl_yz] (tbl_xy) {\(\textsf{edge}(x,y)\)};

  \node[op, above=0.4cm of tbl_xy] (mapyx) {\texttt{Map}};
  \draw[->] (tbl_xy) -- node[labelstyle, xshift=-0.05cm]{$(x, y)$} (mapyx);

  \node[op, above=0.4cm of tbl_yz] (mapzy) {\texttt{Map}};
  \draw[->] (tbl_yz) -- node[labelstyle, xshift=-0.05cm]{$(y, z)$} (mapzy);

  \node[op, above=1.0cm of tbl_rz] (join_map_z) {\texttt{Join-FlatMap} \(z\)};
  \draw[->] (tbl_rz) -- node[labelstyle, xshift=-0.05cm]{$(z)$} (join_map_z);
  \draw[->] (mapzy) -- node[labelstyle]{$(z, y)$} (join_map_z);

  \node[op, above=1.0cm of join_map_z] (join_map_y) {\texttt{Join-FlatMap} \(y\)};
  \draw[->] (join_map_z) -- node[labelstyle, xshift=-0.05cm]{$(y)$} (join_map_y);
  \draw[->] (mapyx) -- node[labelstyle, xshift=0.1cm, yshift=0.01cm]{$(y, x)$} (join_map_y);

  \node[font=\scriptsize, above=0.3cm of join_map_y] (final_reach) {\(\textsf{reach}(x)\)};
  \draw[->] (join_map_y) -- (final_reach);

  
  \node[font=\scriptsize, left=0.05cm of join_map_z.east, yshift=-0.35cm] {\circledwhite{1}};
  \node[font=\scriptsize, left=0.05cm of join_map_y.east, yshift=-0.35cm] {\circledwhite{2}};
  \end{tikzpicture}
  \caption{\small A rooted JST for $r_2$ in Example~\ref{ex:reach} (left) and its translated \ir (right) following a post order traversal of the rooted JST.}
  \label{fig:jst}
\end{figure}
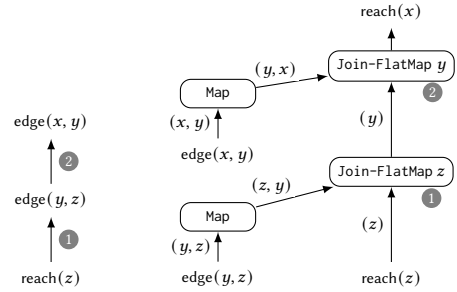

A \textbf{join spanning tree} (JST) extends this notion to cyclic rules. A JST is a maximum spanning tree of the weighted join graph and reduces to a join tree when the rule is acyclic~\cite{Maier83}. A \textbf{rooted JST} defines a join-project plan by post-order traversal: at each step, an atom is joined with its parent, followed by projecting out variables no longer needed. The only JST (also a join graph or a join tree) for Example~\ref{ex:reach} is shown in Figure~\ref{fig:jst}. Here we root the JST at $\textsf{edge}(x, y)$, and it maps into the \ir on the right, i.e. a bottom-up join order. If we root the JST at $\textsf{reach}(z)$ instead, we recover the \ir in Figure~\ref{fig:fused_ir}.

\smallskip
\introparagraph{Search Space} Our search space excludes semijoins (atoms whose variables are subsumed by another), antijoins and filters, as they are pushed down to the lowest possible transformation in the \ir. This narrows down to a multi-way (inner) join where the search space is defined as \textit{all rooted JSTs of its join graph}. There are three reasons for such a choice: (1) JSTs avoid cross products when possible, since cross products correspond to zero-weight edges in the weighted join graph, (2) this space is reasonably small and easy to enumerate~\cite{Winter86}, and (3) it collapses down to rooted join trees for acyclic rules, for which the post-order join-project plans yield tight bounds on the intermediate sizes (proven by~\cite{grounding24}). In our running example (which is acyclic), Figure~\ref{fig:jst} will be selected against Figure~\ref{fig:fused_ir} as it has a cost of 2, instead of 3 in our cost model. Intuitively, it avoids computing $\textsf{edge}(x, y) \bowtie \textsf{edge}(y, z)$ by using two semi-joins.

Next, we show a more involved example and show that JSTs can optimize join-project plans for rules with cyclic multi-way joins.

\begin{figure*}[t]
  \centering
  \begin{minipage}{0.22\linewidth}
    \centering
    \begin{lstlisting}[style=DatalogStyle, mathescape=true, numbers=none, basicstyle=\scriptsize\ttfamily]
VarPointsTo(heap, to) :-
  Reach(inm),
  LoadArrayIdx(base, to, inm),
  VarPointsTo(bh, base),
  ArrayIdxPointsTo(bh, heap),
  VarType(to, tp),
  HeapAllocationType(bh, bht),
  ComponentType(bht, bct),
  SupertypeOf(tp, bct).
  \end{lstlisting}
  \end{minipage}%
  \hfill
  \begin{minipage}{0.36\linewidth}
    \centering
    \begin{tikzpicture}[
        >=latex, 
        auto,
        node distance=0.8cm and 1.2cm,
        op/.style={
          draw,
          rounded corners,
          minimum width=18mm,
          minimum height=4mm,
          font=\ttfamily\tiny,
          align=center
        },
        labelstyle/.style={
          font=\tiny\ttfamily,
          midway,
          fill=white,
          inner sep=0.5pt
        }
    ]

    \node[op] (r2) {\texttt{HeapAllocationType}\\\texttt{(bh, bht)}};
    \node[op, below left=0.5cm and 0.1cm of r2] (r3) {\texttt{VarPointsTo}\\\texttt{(bh, base)}};
    \node[op, below=1.2cm of r3] (r4) {\texttt{LoadArrayIdx} $\bowtie$ \texttt{Reach} \\\texttt{(base, to)}};
    \node[op, below right=0.4cm and 0.1cm of r3] (r5) {\texttt{ComponentType}\\\texttt{(bht, bct)}};
    \node[op, below right=0.5cm and 0.1cm of r2] (r6) {\texttt{ArrayIdxPointsTo}\\\texttt{(bh, heap)}}; 
    \node[op, below=1.1cm of r5] (r7) {\texttt{VarType}\\\texttt{(to, tp)}};
    \node[op, below=1.1cm of r6] (r8) {\texttt{SupertypeOf}\\\texttt{(tp, bct)}};

    \draw[<-, thick] (r2) -- node[labelstyle]{\texttt{bh} \circledwhite{2}} (r3);
    \draw[<-, thick] (r3) -- node[labelstyle, xshift=1mm]{\texttt{base} \circledwhite{1}} (r4);
    \draw[<-, thick] (r2) -- node[labelstyle, yshift=-5mm, xshift=1mm]{\texttt{bht} \circledwhite{5}} (r5);
    \draw[<-, thick] (r2) -- node[labelstyle]{\texttt{bh} \circledwhite{6}} (r6);
    \draw[<-, thick] (r5) -- node[labelstyle]{\texttt{bct} \circledwhite{4}} (r8);
    \draw[->, thick] (r7) -- node[labelstyle]{\texttt{tp} \circledwhite{3}} (r8);
    \draw[dotted] (r3) -- node[labelstyle, xshift=-7mm]{\texttt{bh}} (r6);
    \draw[dotted] (r4) -- node[labelstyle]{\texttt{to}} (r7);
    \end{tikzpicture} 
  \end{minipage}%
  \hfill
  \begin{minipage}{0.42\linewidth}
    \centering
    \includegraphics[width=1.0\linewidth]{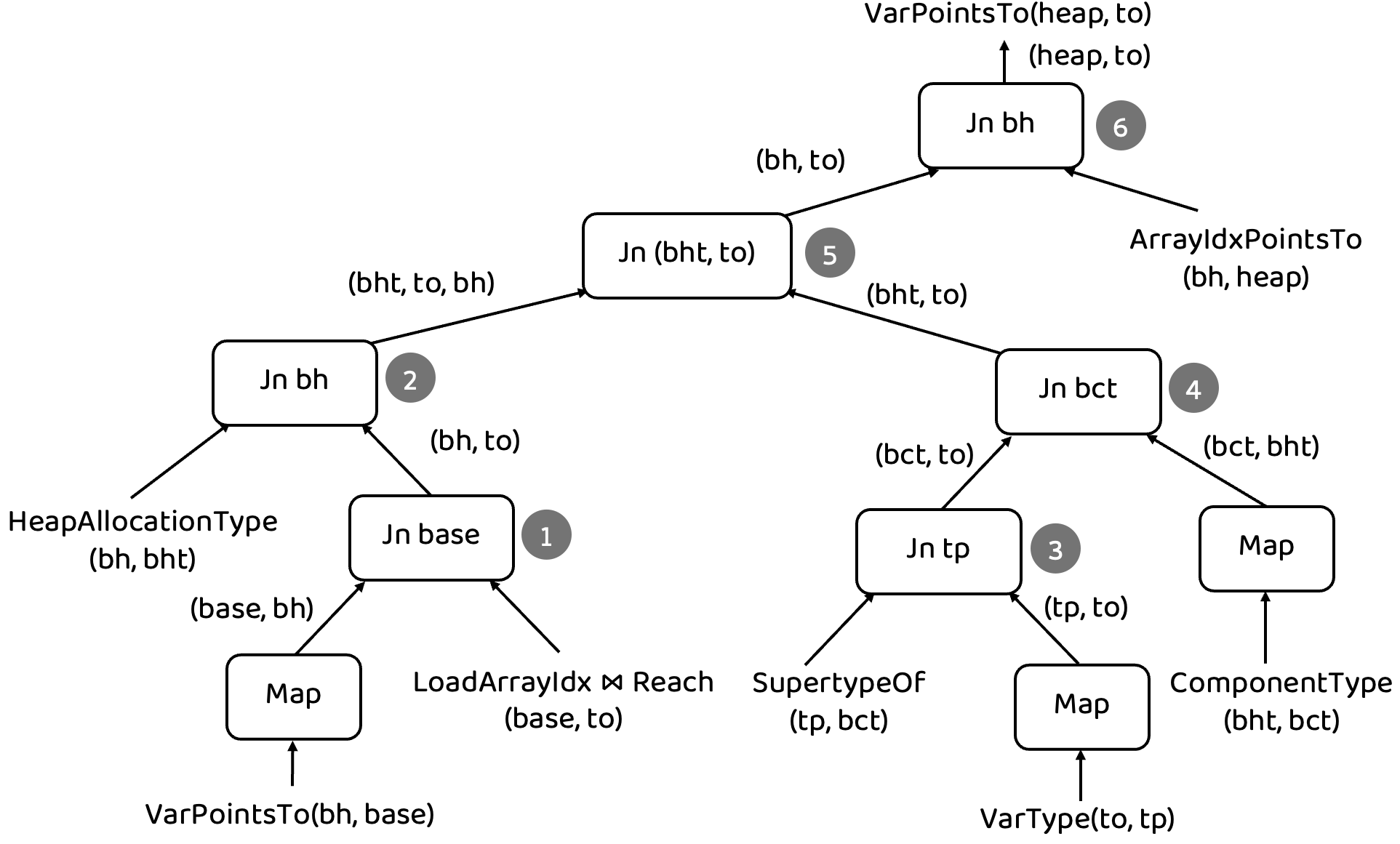}
  \end{minipage}
  \caption{\small The \textsc{doop} rule for Example~\ref{ex:doop} (left); the rooted JST chosen by the optimizer over the cyclic join graph (center, numbered post-order); corresponding \ir following the post-order (right). Semijoin \texttt{Reach(inm)} is pushed to \texttt{LoadArrayIdx}; \texttt{Jn} is a shorthand for \texttt{Join-FlatMap}.}
  \label{fig:doop}
\end{figure*}
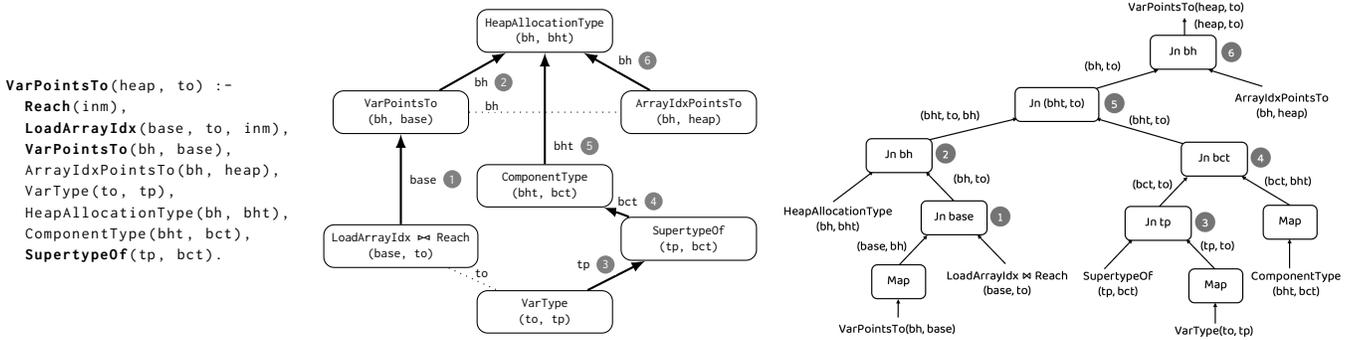

\begin{example} {\label{ex:doop}}
  Consider an expensive rule (Figure~\ref{fig:doop}, left) from the \textsf{DOOP} program analysis framework~\cite{BravenboerS09}, which contains a recursive 8-way join.
  \texttt{VarType}, \texttt{HeapAllocationType}, and \texttt{ComponentType} are \EDBs; others (in bold) are \IDBs. Figure~\ref{fig:doop}(center) shows its join graph. The JST selected by the optimizer is rooted at \texttt{LoadArrayIdx} and is annotated on the join graph by thick, directed edges. The dotted edges are edges in the join graph, but are not part of the rooted JST. Here, \texttt{Reach(inm)} is a semijoin atom to be subsumed by \texttt{LoadArrayIdx} and \texttt{inm} is projected away after semijoin. When creating the \ir, this semijoin is pushed down to the leaf \texttt{LoadArrayIdx}. 
\end{example}

We now discuss query plans for the rule in Example~\ref{ex:doop}. \souffle uses the given listing order: it joins \texttt{Reach} with \texttt{LoadArrayIdx}, then the result joins with \texttt{VarPointsTo}, and so on. The listing order here has been hand-picked for practical efficiency. In our cost model, the costing of this listing order is $5$, dominated by the fifth join with \texttt{HeapAllocationType}: when we finish the first four joins up to \texttt{VarType}, the resulting schema, projecting away unnecessary variables, is $(\texttt{bh}, \texttt{tp}, \texttt{heap}, \texttt{to})$, where $\texttt{bh}$ and $\texttt{tp}$ are join keys for future joins with \texttt{HeapAllocationType} and \texttt{SupertypeOf} respectively; $(\texttt{heap}, \texttt{to})$ are desired output variables. The next join with \texttt{HeapAllocationType} involves 5 variables, that is, \texttt{bh}, \texttt{tp}, \texttt{heap}, \texttt{to} and \texttt{bht}. In contrast, the rooted JST in Figure~\ref{fig:doop}(left) turns into the \ir (Figure~\ref{fig:doop}(right)), \texttt{Jn} being a shorthand for \texttt{Join-FlatMap}. The leaf $\texttt{LoadArrayIdx} \bowtie \texttt{Reach}$ is the semijoin pushdown. The optimizer favors this join-project plan as it has a cost of $3$--no transformation needs more than 3 distinct variables--lower than the listing order.

\subsection{Plan Execution} 
\label{sec:plan_execution}
\smallskip
\introparagraph{Left-deep Plans} Join planning is tied closely to the underlying execution. The listing order \souffle, \ddlog and others used is equivalent to a left-deep join plan in a \dbms and is incapable of representing bushy plans, such as the one in Figure~\ref{fig:doop}. This is a necessary choice for \souffle as it always compiles the listing order into an indexed nested loop join, where indexes (e.g. B-trees) are built on every atom except the first one and it iterates over each tuple of the first atom (as the outer loop) and probes into the indexes of the rest. Left-deep plans allow pipelining without intermediate materialization, which makes \souffle memory-efficient. However, scaling indexed nested loop joins to a multi-threaded setting is non-trivial (e.g.~\cite{LeisBK014}) and modern \Datalog systems such as \souffle and Flan~\cite{Flan24} only distribute the outermost for-loop among threads. As shown in our experiments, this level of parallelism is insufficient to saturate resources even for simple recursive queries like reachability.

\smallskip
\introparagraph{Bushy Plans}
\eclair targets \textsf{DD}, a fundamentally different execution model. Its differential operators are inherently stateful, so intermediate materialization is unavoidable. A key advantage, however, is their asynchrony: changes propagate through the dataflow graph and operators react concurrently without waiting for earlier stages to finish. Hence, \textsf{DD} naturally exploits multicore parallelism, as computation is scheduled dynamically based on available updates rather than a pre-determined execution sequence. Consequently, \eclair can take advantage of bushy plans (i.e. \ir is bushy) without worrying about compromising parallelism. To control memory blow-ups, \eclair leverages insights from database theory (e.g. tree decompositions~\cite{FAQ16,PANDA17,grounding24}, worst-case optimal joins~\cite{NPRR13, freejoin24}, etc.) to select plans that have smallest possible worst-case intermediate sizes. When multiple candidate plans have the same estimated cost, \eclair heuristically prefers bushier (shallower) rooted JSTs.

\section{Making \Datalog Robust}
\label{sec:robust}
\Datalog execution typically exhibits high sensitivity to data distribution, recursive control, join orders, etc. The volatility makes conventional optimization techniques for \texttt{SQL} less effective, or even inapplicable. When a suboptimal plan is chosen for a recursive rule, the iterative nature of \Datalog may exacerbate the slowdown. In contrast, there has been a growing interest in techniques that makes \texttt{SQL} queries robust, such as sideways information passing (\sip)~\cite{IvesT08, LIP17}, predicate transfer~\cite{PT23}, and diamond hardened joins~\cite{DiamondJoin24}. These approaches advocate a more pessimistic stance, prioritizing resilience against worst-case scenarios to ensure stable performance. However, robustness techniques are largely absent in \Datalog systems. Integrating such techniques into \eclair is a first step toward our vision: making \Datalog execution robust. In fact, for a simple $R \bowtie S$, \textsf{DD}'s \textcolor{blue}{\texttt{join}} operator already yields robustness as it runs a symmetric hash-join that balances both inputs (i.e. no build-probe asymmetry)~\cite{arrangement}. For multi-way joins, the optimizer of \eclair uses a \sip-style algorithm (discuss next) to stabilize iterative execution.

\smallskip
\introparagraph{\sip-style Algorithms}
The key idea of \sip-style algorithms is to pre-filter dangling tuples before the actual joins. Dangling tuples are tuples that do not participate in the final output. The Yannakakis algorithm~\cite{Yannakakis81} is a classic example. It requires the join graph to be acyclic to be able to construct a join tree. Following the post-order traversal over the join tree, it applies two sequence of semijoins to pre-filter base tables: (1) a bottom-up pass that semijoins the child atoms with the parent, followed by (2) a top-down pass that uses the reduced parent atoms to further semijoin-reduce the children. The semijoins provably prune all dangling tuples and the subsequent joins exhibit robustness against bad join orders in practice~\cite{LIP17, PT23}.

Many expensive \Datalog rules, however, involve cyclic join graphs, where no join tree exists. Inspired by Yannakakis, we apply a similar two-pass semijoin reduction in \eclair, but for arbitrary join graphs. Our approach is as follows. We pick any atom in the join graph to start a breadth-first search (BFS). As we visit an atom, we semijoin-reduce it using its already-visited neighboring atoms, on the join keys. We conclude the first pass when all atoms are visited. Then we traverse the join graph in the reverse order of the first pass with another round of semijoin reduction (i.e. the second pass).

\begin{example}[Galen] {\label{ex:galen}}
  Consider the following program from~\cite{Galen} that describes an inference task in medical ontologies:
  \begin{lstlisting}[style=DatalogStyle, mathescape=true, numbers=none, basicstyle=\footnotesize\ttfamily]
  r1.  p(x,z) :- p(x,y), p(y,z).
  r2.  p(x,z) :- p(y,w), u(w,r,z), q(x,r,y).
  r3.  p(x,z) :- c(y,w,z), p(x,w), p(x,y).
  r4.  q(x,r,z) :- p(x,y), q(y,r,z).
  r5.  q(x,u,z) :- q(x,r,z), s(r,u).
  r6.  q(x,e,o) :- q(x,y,z), r(y,u,e), q(z,u,o).
  \end{lstlisting}
  where the atoms \texttt{u}, \texttt{c}, and \texttt{s} are \EDBs. The listing orders here are hand-picked for performance on the given dataset~\cite{Galen}. Among them, $r_2, r_3, r_6$ dominate the runtime and they are highly susceptible to bad join orders. The plan optimization in Sec.~\ref{sec:planning} can effectively handle $r_2$--it avoids joining $\texttt{u}$ and $\texttt{q}$ upfront (due to a high cost of $6$ in our cost model). However, the optimal join orders for $r_3$ and $r_6$ are obscure (note that 
  $p, q$ are \IDBs in a mutual recursion). Take $r_3$ for example, it has a triangular join graph and all join orders are indistinguishable under our cost model (i.e. cost of $4$) and yet the chosen order is an order of magnitude faster than, say, \texttt{c(y,w,z), p(x,y), p(x,w)} (we will call it the bad listing order henceforth).
\end{example}
We next show how \sip is applied to $r_3$ (Example~\ref{ex:galen}). We implement it via rule rewriting: we pick \texttt{c} as the start and visits the atoms in the bad listing order. Then the rewritten \sip rules for $r_3$ are the following (underscores are placeholders for unused variables):
\begin{lstlisting}[style=DatalogStyle, mathescape=true, numbers=none, basicstyle=\footnotesize\ttfamily]
  // (1) first pass c(y,w,z) -> p(x,w) -> p(x,y)
  p1(x,y) :- c(y,_,_), p(x,y).
  p2(x,w) :- c(_,w,_), p1(x,_), p(x,w).

  // (2) 2nd pass c(y,w,z) <- p1(x,w) <- p2(x,y)
  p3(x,y) :- p1(x,y), p2(x,_).
  c4(y,w,z) :- c(y,w,z), p2(_,w), p3(_,y).

  // (3) reduced join equivalent to the original r3
  r3'.  p(x,z) :- c4(y,w,z), p3(x,y), p2(x,w).
\end{lstlisting}
The rewriting is equivalent to $r_3$, but it is more robust against poor join orders. Indeed, the bad listing order incurs a substantial output blow-up when joining \texttt{c} and \texttt{p(x,y)}, but the final output shrinks significantly at the last join with \texttt{p(x,w)}. The rewriting, without such runtime knowledge, passes the selective semijoins of \texttt{p2(x,w)} to the first two atoms and reduces them into \texttt{p3}, \texttt{c4} in the second pass. Even though $r_3^{\prime}$ uses the same join order as $r_3$, $\texttt{c4} \bowtie \texttt{p3}$ is now drastically smaller. Similar techniques can be applied to $r_6$. 


\smallskip
\introparagraph{\sip Overheads} The \sip rewriting introduces new intermediate (semijoin) rules, hence new \ir(s): in this example, \texttt{p1}, \texttt{p2}, \texttt{p3}, \texttt{c4}, pre-filter the input tables for \texttt{r3}. This inevitably incurs overheads like maintaining these extra semijoin \ir(s). However, this is outweighed if most of the dangling tuples are pruned. We will show in Sec.~\ref{sec:planvar} that \sip often leaves the structural optimizer (Sec.~\ref{sec:planning}) with a near-worst-case scenario (i.e. all remaining tuples join), and by design, the optimizer strives for worst case optimalities. Future optimizations include incorporating techniques such as predicate transfer~\cite{PT23} to further mitigate the semijoin costs using lightweight Bloom filters.

\section{Subplan Sharing}
\label{sec:sharing}
McSherry et al.~\cite{arrangement} introduced arrangements for \textsf{DD} to enable efficient index sharing across concurrent queries without redundant reconstruction. They demonstrated the benefits on simple \Datalog programs by manually enforcing arrangement sharing.

In \eclair, we extend this idea by automatically detecting and sharing common subplans both within and across rules to reduce memory consumption during execution. The sharing algorithm we use is greedy, exploiting the fact that the executor (and \textsf{DD}) will incrementally maintain the output of every intermediate operator. The input we have is a set of \ir, one for each rule, where each \ir is a logical plan with no sharing components yet. To identify reusable subplans, we normalize each \ir in a canonical form and hash every subtree. If a duplicate hash is detected, the corresponding subtree is truncated and replaced by a pointer to the output of the first occurrence--a shared subplan possibly from the same or a different \ir. We repeat this truncation until no more sharing can be found.

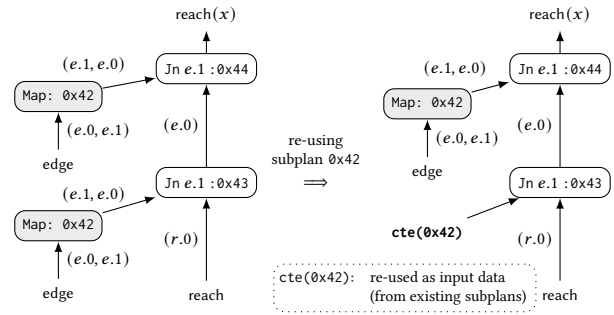
\begin{figure}[t]
  \centering
  \begin{tikzpicture}[
      >=latex, 
      auto,
      node distance=1.2cm and 2cm,
      op/.style={
        draw,
        rounded corners,
        minimum width=11mm,
        minimum height=4mm,
        font=\scriptsize
      },
      sharedop/.style={
        draw,
        rounded corners,
        minimum width=11mm,
        minimum height=4mm,
        font=\scriptsize,
        fill=lightgray!30
      },
      labelstyle/.style={
        font=\scriptsize,
        midway,
        fill=white,
        inner sep=1pt
      },
      dottedbox/.style={
        draw,
        dotted,
        inner sep=4pt,
        rounded corners
      }
  ]

  \node[font=\scriptsize] (tbl_rz) {\(\textsf{reach}\)};
  \node[font=\scriptsize, left=1.3cm of tbl_rz] (tbl_yz) {\(\textsf{edge}\)};
  \node[font=\scriptsize, above=1.3cm of tbl_yz] (tbl_xy) {\(\textsf{edge}\)};

  \node[sharedop, above=0.5cm of tbl_xy] (mapyx) {\texttt{Map: 0x42}};
  \draw[->] (tbl_xy) -- node[labelstyle, xshift=1.0cm]{$(e.0, e.1)$} (mapyx);

  \node[sharedop, above=0.5cm of tbl_yz] (mapzy) {\texttt{Map: 0x42}};
  \draw[->] (tbl_yz) -- node[labelstyle, xshift=1.0cm]{$(e.0, e.1)$} (mapzy);

  \node[op, above=1.1cm of tbl_rz] (join_map_z) {\texttt{Jn} \(e.1\) \texttt{:0x43}};
  \draw[->] (tbl_rz) -- node[labelstyle, xshift=-0.05cm]{$(r.0)$} (join_map_z);
  \draw[->] (mapzy) -- node[labelstyle]{$(e.1, e.0)$} (join_map_z);

  \node[op, above=1.1cm of join_map_z] (join_map_y) {\texttt{Jn} \(e.1\) \texttt{:0x44}};
  \draw[->] (join_map_z) -- node[labelstyle, xshift=-0.05cm]{$(e.0)$} (join_map_y);
  \draw[->] (mapyx) -- node[labelstyle, yshift=0.1cm]{$(e.1, e.0)$} (join_map_y);

  \node[font=\scriptsize, above=0.3cm of join_map_y] (final_reach_left) {\(\textsf{reach}(x)\)};
  \draw[->] (join_map_y) -- (final_reach_left);


  \node[font=\scriptsize, right=0.5cm of join_map_z] (arrow) {\(\Longrightarrow\)};
  \node[font=\scriptsize, above=-0.08cm of arrow, text width=2cm, align=center] (arrow_label) {re-using\\subplan \texttt{0x42}};

  \node[font=\scriptsize, right=4cm of tbl_rz] (shared_rz) {\(\textsf{reach}\)};
  \node[font=\scriptsize, left=1.3cm of shared_rz] (shared_yz) {};
  \node[font=\scriptsize, above=1.3cm of shared_yz] (shared_xy) {\(\textsf{edge}\)};

  \node[sharedop, above=0.5cm of shared_xy] (shared_mapyx) {\texttt{Map: 0x42}};
  \draw[->] (shared_xy) -- node[labelstyle, xshift=1.0cm]{$(e.0, e.1)$} (shared_mapyx);

  \node[font=\scriptsize, above=0.5cm of shared_yz] (hash_value) {\textbf{\texttt{cte(0x42)}}};

  \node[op, above=1.1cm of shared_rz] (shared_join_map_z) {\texttt{Jn} \(e.1\) \texttt{:0x43}};
  \draw[->] (shared_rz) -- node[labelstyle, xshift=-0.05cm]{$(r.0)$} (shared_join_map_z);
  \draw[->] (hash_value) -- node[labelstyle]{} (shared_join_map_z);

  \node[op, above=1.1cm of shared_join_map_z] (shared_join_map_y) {\texttt{Jn} \(e.1\) \texttt{:0x44}};
  \draw[->] (shared_join_map_z) -- node[labelstyle, xshift=-0.05cm]{$(e.0)$} (shared_join_map_y);
  \draw[->] (shared_mapyx) -- node[labelstyle, yshift=0.1cm]{$(e.1, e.0)$} (shared_join_map_y);

  \node[font=\scriptsize, above=0.3cm of shared_join_map_y] (final_reach_right) {\(\textsf{reach}(x)\)};
  \draw[->] (shared_join_map_y) -- (final_reach_right);


  \node (legend) [anchor=north, font=\scriptsize, align=left,
      inner sep=2pt, yshift=-2mm, xshift=-3mm,
      draw, dotted, rounded corners, fill=white]
  at (hash_value.south) {
  \begin{tabular}{@{}l@{\quad}l@{}}
      \texttt{cte(0x42)}: & re-used as input data  \\
      & (from existing subplans) \\
  \end{tabular}
  };

  \end{tikzpicture}
  \caption{\small Subplan sharing within \ir of Fig.~\ref{fig:jst} (reused \mapop are shaded)}
  \label{fig:shared_ir}
\end{figure}

Figure~\ref{fig:shared_ir} illustrates this process on the \ir in Figure~\ref{fig:jst}. The left simply canonicalizes this \ir by encoding variable positions relative to their atoms. For example, in $\textsf{edge}(x, y)$ and $\textsf{edge}(y, z)$, variables are rewritten as $(e.0, e.1)$, while $\textsf{reach}$ maps to $r.0$. This reveals that two \mapop subplans are identical (up to variable renaming, having the same hash value, say \texttt{0x42}. Thus, we reuse the output of the first \mapop, i.e. an index on key $e.1$ and value $e.0$, for the second instance (as the leaf node). Figure~\ref{fig:shared_ir}(right) shows the resulting \ir after sharing. 

Our approach subsumes shared arrangements~\cite{arrangement} and further extends it to sharing common table subexpressions (CTEs), e.g. if there is another \ir requesting the first column $e.0$ from $\textsf{edge}(x, y) \bowtie \textsf{reach}(y)$, the optimizer will not re-construct, but instead directly link to the output of the lower \texttt{Join-FlatMap} of Figure~\ref{fig:shared_ir}, say, \texttt{0x43}. 


\smallskip
\introparagraph{Summary} Figure~\ref{fig:shared_ir}(right) is the final \ir for $r_2$ (Example~\ref{ex:reach}) that \eclair's optimizer sends to its executor. To summarize our optimizations (of each rule), we \circledwhite{1} enumerate every rooted JST of the rule, chooses the best based on the cost model (Sec.~\ref{sec:planning}), and maps it into an \ir, \circledwhite{2} create new \ir(s) for auxiliary semijoins (or \sip rules) to pre-filter atoms as reduced inputs for the original \ir (Sec.~\ref{sec:robust}). Along the way, we \circledwhite{3} aggressively fuse operators (Sec.~\ref{sec:fusion}), and \circledwhite{4} hash every subplan to maximize CTE reuse within and across \ir.


\section{Boolean Specialization} \label{sec:specialization}

This section introduces a novel optimization of \eclair's executor. For each row $(\texttt{data}, \texttt{time}, \texttt{diff})$, \textsf{DD} uses an integral (e.g. 64 bits) \texttt{diff} to encode the number of copies of \texttt{data}, 0 being its absence and negative values indicating deletions or subtractions. Differential operators use integer arithmetics (e.g. $+, \cdot, -$) to track output \texttt{diff}s. A $\textcolor{blue}{\texttt{join}}$ multiplies \texttt{diff}s of the two matching rows; $\textcolor{blue}{\texttt{concat}}$ of $((\textsf{a}, \textsf{b}), 0, 4)$ and $((\textsf{a}, \textsf{b}), 0, 3)$ results in $((\textsf{a}, \textsf{b}), 0, 7)$ by adding the \texttt{diff}s, while $\textcolor{blue}{\texttt{antijoin}}$ yields $((\textsf{a}, \textsf{b}), 0, 1)$ by subtracting the \texttt{diff}. 

Integer arithmetic on \texttt{diff} fits well for incremental execution, but it is not always necessary for \Datalog, where the mere presence of a tuple may be enough. In \textsf{DD}, it corresponds to restricting the \texttt{diff} to the Booleans, i.e. \texttt{true} for presence and \texttt{false} otherwise. Then, $\textcolor{blue}{\texttt{join}}$ implies a logical \textsf{AND}, as the output row is present only if both inputs are. A $\textcolor{blue}{\texttt{concat}}$ of $((\textsf{a}, \textsf{b}), 0, \texttt{true})$ and $((\textsf{a}, \textsf{b}), 0, \texttt{false})$ returns $((\textsf{a}, \textsf{b}), 0, \texttt{true} \vee \texttt{false} = \texttt{true})$ by a logical \textsf{OR}—present when at least one exists. However, an $\textcolor{blue}{\texttt{antijoin}}$ is not expressible because subtraction is undefined for Booleans. In fact, \textsf{DD} encodes \textit{presence} of a tuple as a zero-bit struct; a \texttt{false} (or \textit{absent}) tuple is dropped upon encounter. A Boolean \texttt{diff} has two benefits: (1) it reduces memory footprint by storing \texttt{diff} as a zero-bit presence struct, and (2) the logical simplifications enable compiler optimizations that short-circuit Boolean operations.

As such, \eclair enforces Boolean \texttt{diffs} (as the zero-bit presence struct) whenever possible by coercing the input \texttt{diff} values into Booleans and each differential operator, such as \textcolor{blue}{\texttt{join}}, \textcolor{blue}{\texttt{concat}}, into Boolean semantics, i.e. \textsf{AND}, \textsf{OR} operations. To handle non-monotonic operators (i.e. those that require negation or deletion operations), e.g. antijoins, we introduce a custom \textcolor{blue}{\texttt{lift}} operator that transitions \texttt{diff} values between any data types, e.g. lift $\texttt{true}$ to $1$ and $\texttt{false}$ to $0$. In our example, $((\textsf{a}, \textsf{b}), 0, \texttt{true})$ \textcolor{blue}{\texttt{antijoin}} $((\textsf{a}, \textsf{b}), 0, \texttt{true})$ is lifted to an integer subtraction $((\textsf{a}, \textsf{b}), 0, 1-1=0)$, and then casted down to a Boolean \texttt{diff} of \texttt{false}. As a general technique, \eclair also supports \texttt{diff} field over other arithmetics (e.g. \texttt{MIN}) to express and optimize recursive aggregation (see next).

\section{Extensibility}
\label{sec:extensibility}



\smallskip
\introparagraph{Algebraic Semantics} 
A first extension (and optimization) is on the algebraic types for \texttt{diff}. Sec.~\ref{sec:specialization} encodes \texttt{diff} as a Boolean value, to optimize batch \Datalog queries. For {\em incremental \Datalog} (\EDBs have insertions/deletions over time), we fall back on integer arithmetic as \textsf{DD} normally does. Recursive aggregation arises frequently in modern data analytics~\cite{Flix16,WangK0PS22,POPS22,NestedRec24}. For example, this program computes connected components (CC) of an undirected graph:

\begin{lstlisting}[style=DatalogStyle, mathescape=true, numbers=none, basicstyle=\footnotesize\ttfamily]
  CC(x,MIN(x)):- edge(x,_). // initialize labels i
  CC(x,MIN(i)):- edge(y,x), CC(y,i). // propagate MIN(i)
\end{lstlisting}
The program iteratively propagates labels—discarding larger ones as smaller labels are found—until all nodes in a connected component share the same label. Semi-naïve execution can not handle this recursive \texttt{MIN} because prior facts may be retracted. In fact, most \Datalog engines require invasive changes to support this (e.g. Souffle lacks recursive aggregation to-date). In contrast, \eclair needs only minor glue: it reuses \textsf{DD}'s native incremental machinery (cf. \textsf{CC}, \textsf{SSSP} in Table~\ref{tab:results}). Similar to the Boolean specialization (Sec.~\ref{sec:specialization}), \eclair implemented this by baking aggregations directly into the \texttt{diff} field via a \textit{monoid}~\cite{Semirings07}. Intuitively, it is a carrier set plus certain on-top operations; Booleans for batch \Datalog were $({\texttt{bool}}, \vee)$, antijoins or incremental \Datalog use integers $({\mathbb{Z}}, +/-/\cdot)$, while \texttt{CC} uses labels and a \texttt{MIN} operator $({\mathbb{Z}}, \texttt{MIN})$. Selecting a monoid—and, when needed, applying \textcolor{blue}{\texttt{lift}} to cast between monoids—yields efficient, general aggregation semantics without executor redesign.


\smallskip
\introparagraph{Distributed Execution}
A natural extension is scaling \eclair to distributed environments. \textsf{DD} operators support scale-out execution by-design: they are sharded across workers (i.e. threads or machines) to attain resource saturation. Unlike \souffle and \recstep, which are grounded in single-machine execution because of their designs, or \textsf{BigDatalog}~\cite{DBLP:conf/sigmod/ShkapskyYICCZ16}
that carefully re-designs distributed execution for recursive settings, \eclair, as future work, can seamlessly extend beyond single-node \Datalog setups.

\begin{table*}[t]
    \footnotesize
    \centering
    \caption{\small Runtimes (seconds) for 4 and 64 threads (shown as 4|64). Lower is better. Per row, the best performance is colored in \high{blue} for 4 threads and \higher{red} for 64 threads; \texttt{TO} = 900s timeout; \texttt{OOM} = out of memory. Unsupported cases (e.g., \textsf{mutual} or \textsf{nonlinear recursion}) state the reason in-cell. }
    \begin{tabular}{lc||p{0.8cm}>{\columncolor{gray!20}}p{0.8cm}|p{0.8cm}>{\columncolor{gray!20}}p{0.8cm}|p{0.8cm}>{\columncolor{gray!20}}p{0.8cm}|p{0.8cm}>{\columncolor{gray!20}}p{0.8cm}|p{0.8cm}>{\columncolor{gray!20}}p{0.8cm}|p{0.8cm}>{\columncolor{gray!20}}p{0.8cm}}
        \toprule
        \textbf{\textsf{Program}, \#rules} & \textbf{\textsf{Dataset}} & \multicolumn{2}{c|}{\textsf{FlowLog} (4 | 64)} & \multicolumn{2}{c|}{{Souffle (4 | 64)}} & \multicolumn{2}{c|}{{RecStep (4 | 64)}} & \multicolumn{2}{c|}{{DDlog (4 | 64)}} & \multicolumn{2}{c|}{{DuckDB (4 | 64)}} & \multicolumn{2}{c}{{Umbra (4 | 64)}} \\
        \midrule
        \multirow{4}{*}{\textsf{CC}, 2~\cite{recstep}} & livejournal                             & \high{46.0} & \higher{9.1}   & \multicolumn{2}{c|}{\multirow{4}{*}{\thickcross \textsf{ recurs. aggregate}}}     & 90.0 & 28.1    & 196.1 & 116.2 & 121.2 & 96.8 & 88.94 & 26.5 \\
        & orkut                  & 67.9 & \higher{13.5}   & \multicolumn{2}{c|}{}      & 122.5 & 26.6    & 307.7 & 185.7 & 46.2& 29.0 & \high{42.5} & 23.3 \\
        & arabic                                & \high{403.3} & \higher{67.7}                          & \multicolumn{2}{c|}{}           & \texttt{TO} & 252.0            & \texttt{TO} & \texttt{TO} & 632.1 & 397.81 & \texttt{OOM} & \texttt{OOM} \\
        & twitter                               & \texttt{TO} & \higher{145.1}                          & \multicolumn{2}{c|}{}           & \texttt{TO} & 488.1            & \texttt{TO} & \texttt{TO}* & \texttt{TO} & 485.2 & \texttt{OOM} & \texttt{OOM} \\
        \midrule
        \multirow{4}{*}{\textsf{Reach}, 2~\cite{recstep}} & livejournal                      & 11.3 & \higher{5.1}  & 21.5 & 19.1   & 21.3 & 9.1     & 112.3 & 104.1 & \high{6.7} & 6.0 & 13.8 & 11.5 \\
        & orkut                        & 19.1 & 9.2         & 38.3 & 32.9   & 33.5 & 12.6 & 186.7 & 172.9 & \high{11.0} & \higher{8.6} & 21.9 & 18.8 \\
        & arabic                       & 87.7 & 40.8                      & 206.8 & 179.7         & 264.9 & 63.7          & \texttt{TO} & \texttt{TO} & \high{60.4} & \higher{35.0} & 95.0 & 74.8 \\
        & twitter       & 274.8 & 94.8 & \texttt{TO} & \texttt{TO} & 543.5 & 101.7 & \texttt{TO} & \texttt{TO} & \high{121.6} & \higher{62.9} & 189.6 & 167.0 \\
        \midrule
        \multirow{4}{*}{\textsf{SSSP}, 2~\cite{recstep}} & livejournal                      & \high{13.5} & \higher{6.3}                        & \multicolumn{2}{c|}{\multirow{4}{*}{\thickcross \textsf{ recurs. aggregate}}}         & \multicolumn{2}{c|}{\multirow{4}{*}{\thickcross \textsf{ syntax error}}}          & 205.2 & 152.2 & 144.0 & 75.5 & 17.5 & 14.8 \\
        & orkut                        & \high{21.2} & \higher{9.1}                        & \multicolumn{2}{c|}{}          & \multicolumn{2}{c|}{}          & 332.5 & 237.2 & 70.73 & 40.4 & 24.8 & 22.5 \\
        & arabic                       & \high{99.8} & \higher{45.9}                        & \multicolumn{2}{c|}{}          & \multicolumn{2}{c|}{}          & \texttt{TO} & \texttt{TO} & \texttt{TO} & \texttt{TO} & 139.5 & 118.58 \\
        & twitter                      & 302.9 & \higher{105.1}                        & \multicolumn{2}{c|}{}          & \multicolumn{2}{c|}{}          & \texttt{TO} & \texttt{TO} & \texttt{TO} & \texttt{TO} & \high{250.0} & 222.89 \\
        \midrule
        \multirow{3}{*}{\textsf{TC}, 2~\cite{recstep}} & \textsf{G10K-0.001} & 78.2 & 8.5        & 112.7 & 39.7 & 127.0 & 69.5   & 209.2 & 106.4 & 75.2 & 78.5 & \high{23.1} & \higher{6.1} \\
        & \textsf{G20K-0.001} & \high{542.8} & \higher{42.4}                       & 629.6 & 249.2             & \textsf{703.7} & 282.7          & \texttt{TO} & 476.1 & 717.4 & 712.7 & \texttt{OOM} & \texttt{OOM} \\
        & \textsf{G40K-0.001}  & \texttt{TO} & \higher{305.7}        & \texttt{TO} & 668.5   & \texttt{TO} & \texttt{TO}   & \texttt{TO} & \texttt{TO} & \texttt{TO} & \texttt{TO} & \texttt{OOM} & \texttt{OOM} \\
        \midrule
        \multirow{3}{*}{\textsf{SG}, 2~\cite{recstep}} & \textsf{G10K-0.001} & \high{177.6} & \higher{18.6} & 379.0 & 41.3 & 427.5 & 161.8 & 361.9 & 122.0 & 674.4 & 670.8 & \texttt{OOM} & \texttt{OOM} \\
        & \textsf{G20K-0.001} & \texttt{TO} & \higher{90.7}                        & \texttt{TO} & 404.5            & \texttt{TO} & \texttt{TO}          & \texttt{TO} & 575.4 & \texttt{TO} & \texttt{TO} & \texttt{OOM} & \texttt{OOM} \\
        & \textsf{G40K-0.001}  & \texttt{TO} & \higher{815.9}        & \texttt{TO} & \texttt{TO}   & \texttt{TO} & \texttt{TO}   & \texttt{TO} & \texttt{TO} & \texttt{TO} & \texttt{TO} & \texttt{OOM} & \texttt{OOM} \\
        \midrule
        \multirow{3}{*}{\textsf{Bipartite}, 4~\cite{yang2022efficient}} & netflix & 27.8 & 9.3 & 107.6 & 103.8 & 34.8 & \higher{8.0}     & 174.0 & 153.2 & \high{12.1} & 10.62 & 19.3 & 16.83 \\
        & roadca                           & \high{3.5} & \higher{1.3}                       & 29.5 & 6.0         & 145.2 & 172.5          & 15.2 & 13.0 & 22.9 & 11.1 & 42.6 & 33.3 \\
        & mag                              & 306.7 & \higher{76.7}                       & 393.9 & 306.2         & 489.4 & 103.6          & \texttt{TO} & \texttt{TO} & \high{167.9} & 175.5 & 192.7 & 171.5 \\
        \midrule
        \multirow{3}{*}{\textsf{CSDA}, 2~\cite{recstep}} & httpd                             & 4.1 & \higher{1.3}   & 15.2 & 9.8      & 56.1 & 45.5    & 24.0 & 22.0 & \high{3.9} & 6.6 & 18.16 & 7.6 \\
        & linux                             & \high{22.5} & \higher{6.4}  & 145.2 & 77.4    & 599.7 & 272.9 & 121.5 & 109.1 & 41.7 & 23.5 & 94.3 & 27.4 \\
        & postgresql          & \high{11.4} & \higher{4.9}  & 125.1 & 40.4    & 341.5 & 206.3  & 73.9 & 69.2 & 25.3 & 10.1 & 83.0 & 23.0 \\
        \midrule
        \multirow{3}{*}{\textsf{CSPA}, 2~\cite{recstep}} & httpd                             & 112.6 & \higher{14.4}       & \high{67.8} & 50.9 & 382.4 & 154.3 & 319.3 & 290.3 & \multicolumn{2}{c|}{\multirow{3}{*}{\thickcross \textsf{ mutual, nonlin.}}} & \multicolumn{2}{c}{\multirow{3}{*}{\thickcross \textsf{ mutual, nonlin.}}} \\
        & linux                             & 25.1 & \higher{4.8}  & \high{20.9} & 14.3    & 74.8 & 54.5    & 71.2 & 55.1 & \multicolumn{2}{c|}{} & \multicolumn{2}{c}{} \\
        & postgresql          & 120.4 & \higher{15.0}       & \high{76.7} & 56.1 & 344.6 & 161.3 & 326.7 & 282.0 & \multicolumn{2}{c|}{} & \multicolumn{2}{c}{} \\
        \midrule
        \multirow{2}{*}{\textsf{Andersen}, 4~\cite{recstep}} & medium                   & \high{7.9} & \higher{4.2}   & 85.1 & 31.1     & 32.7 & 12.0    & 52.0 & 50.5 & \multicolumn{2}{c|}{\multirow{2}{*}{\thickcross \textsf{ nonlinear recurs.}}} & \multicolumn{2}{c}{\multirow{2}{*}{\thickcross \textsf{ nonlinear recurs.}}} \\
        & large & \high{16.0} & \higher{5.7}   & 187.8 & 69.6     & 63.7 & 17.4    & 107.4 & 103.0 & \multicolumn{2}{c|}{} & \multicolumn{2}{c}{} \\
        \midrule
        \multirow{2}{*}{\textsf{Dyck}, 7} & kernel                             & \high{5.0} & \higher{1.4}   & 17.7 & 10.3     & 21.5 & 14.0    & 27.3 & 27.7 & \multicolumn{2}{c|}{\multirow{2}{*}{\thickcross \textsf{ nonlinear recurs.}}} & \multicolumn{2}{c}{\multirow{2}{*}{\thickcross \textsf{ nonlinear recurs.}}} \\
        & postgresql                            & \high{3.8} & \higher{0.9}   & 12.1 & 6.1     & 15.4 & 10.2    & 15.4 & 14.7 & \multicolumn{2}{c|}{} & \multicolumn{2}{c}{} \\
        \midrule
        \textsf{Galen}, 8~\cite{David2021} & galen             & \high{32.2} & \higher{8.7}  & 59.3 & 36.8    & 486.5 & 667.9      & 111.6 & 64.6 & \multicolumn{2}{c|}{\thickcross \textsf{ mutual, nonlin.}} & \multicolumn{2}{c}{\thickcross \textsf{ mutual, nonlin.}} \\
        \midrule
        \textsf{CRDT}, 23 & crdt                               & 248.3 & \higher{62.3}      & 177.7 & 230.2 & \multicolumn{2}{c|}{\thickcross \textsf{ syntax error}} & \texttt{TO} & 482.1 & \texttt{TO} & \texttt{TO} & \high{58.9} & 132.8 \\
        \midrule
        \textsf{Polonius}, 37 & polonius                           & 215.4 & \higher{41.4}       & 202.4 & 337.9 & \multicolumn{2}{c|}{\thickcross \textsf{ syntax error}} & 583.1 & 526.4 & \texttt{TO} & \texttt{TO} & \high{70.5} & 67.1 \\
        \midrule
        \multirow{2}{*}{\textsf{DDISASM}, 28~\cite{cvc5,Z3}} & cvc5                   & 87.5 & \higher{12.6}        & \high{27.3} & 14.5  & \multicolumn{2}{c|}{\multirow{2}{*}{\thickcross \textsf{ syntax error}}} & 438.6 & 111.3 & \multicolumn{2}{c|}{\multirow{2}{*}{\thickcross \textsf{ mutual, nonlin.}}} & \multicolumn{2}{c}{\multirow{2}{*}{\thickcross \textsf{ mutual, nonlin.}}} \\
        & z3                       & \high{106.0} & \higher{27.9} & 125.9 & 109.9  & \multicolumn{2}{c|}{} & 769.2 & 510.6 & \multicolumn{2}{c|}{} & \multicolumn{2}{c}{} \\
        \midrule
        \multirow{5}{*}{\textsf{DOOP}, 136~\cite{DaCapo}} & batik                              & \high{65.2} & \higher{22.9}        & 651.1 & 160.2   & \multicolumn{2}{c|}{\multirow{5}{*}{\thickcross \textsf{ syntax error}}} & 151.4 & 126.6 & \multicolumn{2}{c|}{\multirow{5}{*}{\thickcross \textsf{ mutual, nonlin.}}} & \multicolumn{2}{c}{\multirow{5}{*}{\thickcross \textsf{ mutual, nonlin.}}} \\
        & biojava                            & \high{10.7} & \higher{7.7}   & 310.4 & 71.9    & \multicolumn{2}{c|}{} & 71.2 & 39.5 & \multicolumn{2}{c|}{} & \multicolumn{2}{c}{} \\
        & eclipse                            & \high{50.1} & \higher{18.2}        & 279.3 & 106.3   & \multicolumn{2}{c|}{} & 139.1 & 118.9 & \multicolumn{2}{c|}{} & \multicolumn{2}{c}{} \\
        & xalan                              & \high{6.9} & \higher{6.3}   & 78.2 & 26.3      & \multicolumn{2}{c|}{} & 49.4 & 29.2 & \multicolumn{2}{c|}{} & \multicolumn{2}{c}{} \\
        & zxing                & \high{10.5} & \higher{8.4}   & 89.7 & 31.9     & \multicolumn{2}{c|}{} & 57.5 & 27.5 & \multicolumn{2}{c|}{} & \multicolumn{2}{c}{} \\
        \bottomrule
    \end{tabular}
    \label{tab:results}
\end{table*}

\section{Benchmarking and Experiments} \label{sec:experiments}


\smallskip
\introparagraph{Programs and Datasets} We curate a benchmark suite that, to our knowledge, is the broadest yet for modern \Datalog engines. It subsumes nearly all publicly available programs and datasets used in recent publications of \recstep, \souffle, and \ddlog~\cite{recstep,David2021,souffleInterp,DBLP:conf/lopstr/ArchHZSS22,dyck22}, plus several new programs/datasets we created or harvested from popular open-source projects. The suite spans across graph analytics, business intelligence, and static program analysis; and stresses diverse recursion behaviors (multi-way joins, mutual, nonlinear, deep iterations, aggregations/antijoins, etc.): (1) \textsf{Bipartite} is a custom program that decides if a connected undirected graph is bipartite (i.e., two-colorable), starting from an initialized blue node:
\begin{lstlisting}[style=DatalogStyle, mathescape=true, numbers=none, basicstyle=\footnotesize\ttfamily]
    red(y)  :- edge(x, y), blue(x). 
    blue(y) :- edge(x, y), red(x).
    answer() :- red(x), blue(x).
\end{lstlisting}
(2) Graph queries: single-source \textsf{Reachability (Reach)}, \textsf{Shortest Path} (\textsf{SSSP}), \textsf{Same Generation (SG)}, \textsf{Transitive Closure (TC)}, and \textsf{Connected Components (CC)}; (3) Program analysis: \textsf{Andersen}, \textsf{CSPA}, and \textsf{CSDA} (Context-sensitive Point-to and Dataflow Analysis) from \recstep's suite~\cite{recstep} (so are the datasets);
(3) \textsf{Dyck} represents \textsf{Dyck}-2 reachability~\cite{dyck22} and uses the two largest CFPQ instances (kernel, postgre)~\cite{dyck22}\footnote{\url{https://formallanguageconstrainedpathquerying.github.io/CFPQ_Data}};
(4) \textsf{Galen} (Example~\ref{ex:galen} for medical ontologies), and \textsf{CRDT} (conflict-free replicated data types) come from McSherry's blog\footnote{\url{https://github.com/frankmcsherry/dynamic-datalog}};
(5) \textsf{Polonius} (alias analysis) for Rust’s borrow checker from an open-sourced project\footnote{\url{https://github.com/rust-lang/polonius}};
(6) \textsf{DOOP}~\cite{BravenboerS09}, a popular Java analysis framework, featureing 136 rules with complicated recursions. Similar to \souffle and Flan's evaluation, the datasets are sampled among the largest from the DaCapo suite~\cite{DaCapo};
(7) \textsf{DDISASM} is a simplified disassembly analysis program from~\cite{DDISASM} and its datasets are synthesized from the widely used SMT solvers: CVC5~\cite{cvc5} and Z3~\cite{Z3}.
For all benchmarks, we use integer data type for inputs; string-valued attributes are pre-hashed to integers before \Datalog execution.

\smallskip
\introparagraph{Competing Engines} We compare against several state-of-the-art, open-source Datalog engines: (1) \textbf{\souffle} (compiled) \footnote{Soufflé interpreter~\cite{souffleInterp} is in general 1.5× slower and hence excluded from the paper.} with its mature optimizer~\cite{souffleInterp, IndexSouffle, TrieSouffle, DBLP:conf/lopstr/ArchHZSS22}; (2) \textbf{\recstep}~\cite{recstep}, which attains stronger performance and scalability over similar designs such as \textsf{BigDatalog}; (3) \textbf{\ddlog}~\cite{ddlog}, which shares the same \textsf{DD} backend as \eclair but differs in design and lacks key optimizations (e.g. memory reduction) discussed in this paper, making it a fair baseline. We also report results from (4) \textbf{DuckDB}~\cite{DuckDB}, and (5) \textbf{Umbra}~\cite{DiamondJoin24}—two highly optimized, state-of-the-art database engines that can execute a subset of our \Datalog workloads in \texttt{SQL}. DuckDB has very recently added \texttt{USING} \texttt{KEY} optimizations~\cite{UsingKEY} for recursive CTEs, and Umbra implements worst-case optimal join~\cite{UmbraWCOJ} and employs query compilation which allows advanced loop compilations for recursive queries~\cite{SichertN22}. However, \texttt{SQL} has limited recursion syntax, and as~\cite{DBLP:journals/corr/abs-2504-02443} and Table~\ref{tab:results} show, nearly half of our benchmarks cannot be directly executed on both databases due to unsupported mutual or nonlinear recursion. We thus exclude these cases from our evaluation.

\souffle (compiled) and \ddlog require per-program compilation, whereas others (e.g. \eclair) are interpreters. \souffle incurs a $\sim$10s compilation for small programs and 30s for larger ones such as \textsf{DOOP}. \ddlog exhibits much longer Rust compilation~\cite{souffleInterp}—often >100s—even for small programs, due to heavy \textsf{DD} dependencies.

\smallskip
\introparagraph{Environment Setup} We evaluate all engines in their latest releases on runtime, scalability, and memory usage. Experiments run on a CloudLab virtual machine~\cite{cloudlab} with two AMD EPYC 7543 32-core processors (64 physical cores, hyper-threading), running Ubuntu 22.04 LTS with 256 GB RAM.

\subsection{Runtime Summary}
Table~\ref{tab:results} summarizes runtimes (in seconds) of all engines across our benchmark, using 4 and 64 threads. To account for join-order effects among \Datalog engines, for each program-dataset pair we construct up to five plausible, semantically equivalent join-order variants (avoiding cross products); when fewer than five exist, we run all possibilities. We run every variant and report median runtimes for each cell. For DuckDB and Umbra, we use semantically equivalent \texttt{SQL}; when multiple \texttt{WITH} \texttt{RECURSIVE} formulations are possible (e.g., \texttt{USING} \texttt{KEY} for \textsf{CC} in DuckDB), we take the best-performing one and report the median over five runs. The fastest per row is marked in {\bf bold} (\textcolor{blue}{\bf blue} for 4 threads, \textcolor{red}{\bf red} for 64 threads).
As discussed in Sec.~\ref{sec:planning}, each join-order variant is executed as a left-to-right binary-join plan in \souffle and \ddlog. \recstep reoptimizes join orders on the fly via its underlying \dbms optimizer; DuckDB/Umbra rely on their own optimizers for recursive CTEs. All \eclair optimizations discussed in prior sections (Sec~\ref{sec:fusion}-\ref{sec:extensibility}) are applied (including structural optimizations and \sip), as a contrast to \ddlog, being a baseline that directly translates \Datalog into \textsf{DD} programs.

On \underline{4 threads}, \eclair outperforms all competitors in \underline{21 out of 41} program-dataset pairs. It consistently leads on programs such as \textsf{Andersen}, \textsf{Dyck}, and \textsf{DOOP}. For example, on \textsf{Andersen} (large), \eclair runs 11.7× faster than \souffle, 4.0× than \recstep, and 6.7× than \ddlog. This advantage comes largely from logic fusion (Sec.~\ref{sec:fusion}) and subplan reuse (Sec.~\ref{sec:sharing}): necessary indexes for \IDBs are built once and reused in multiple rules, avoiding a large amount of redundant work. That said, \eclair does not always excel on \textbf{batch-oriented workloads}—programs having few but expensive iterations, such as \textsf{Reach} and \textsf{CSPA}. In such cases, its incrementality incurs overheads, as large intermediate results are maintained but used sparingly. \textsf{Reach} (twitter, 12 iterations) involves a single expensive join where DuckDB and Umbra's highly optimized vectorized hash join implementations outperform \eclair (\souffle and \ddlog timed out). Similarly, on \textsf{CSPA} (\texttt{httpd}, 29 iterations), \souffle achieves the best performance at 67.8s—1.6× faster than \eclair's 112s runtime.

However, \eclair exhibits markedly superior scalability compared to others. At \underline{64 threads}, it shows substantial speedups over all other engines and emerges as the fastest over \underline{36 out of 41} cases! Even in previously disadvantageous workloads like \textsf{CSPA}, \eclair demonstrates exceptional scaling (e.g., 7.8× speedup on \texttt{httpd}), while \souffle sees only modest gains (e.g., 1.3×)—becoming 3.5× slower than \eclair. This scalability advantage is consistently observed across workloads: on \textsf{Polonius}, Umbra's runtime decreases marginally from 70.5 to 67.1 as thread count increases, \eclair cuts down 215s to 41.4s. On \textsf{DDISASM} (\texttt{cvc5}), \eclair transforms an initial 3.2× slowdown against \souffle (at 4 threads) into a 1.2× speedup. \textsf{Reach} represents the sole exception where DuckDB maintains its lead; however, this advantage stems largely from optimized data loading rather than core execution performance (discuss next).

\begin{figure}[t]
  \centering
  \includegraphics[width=\columnwidth]{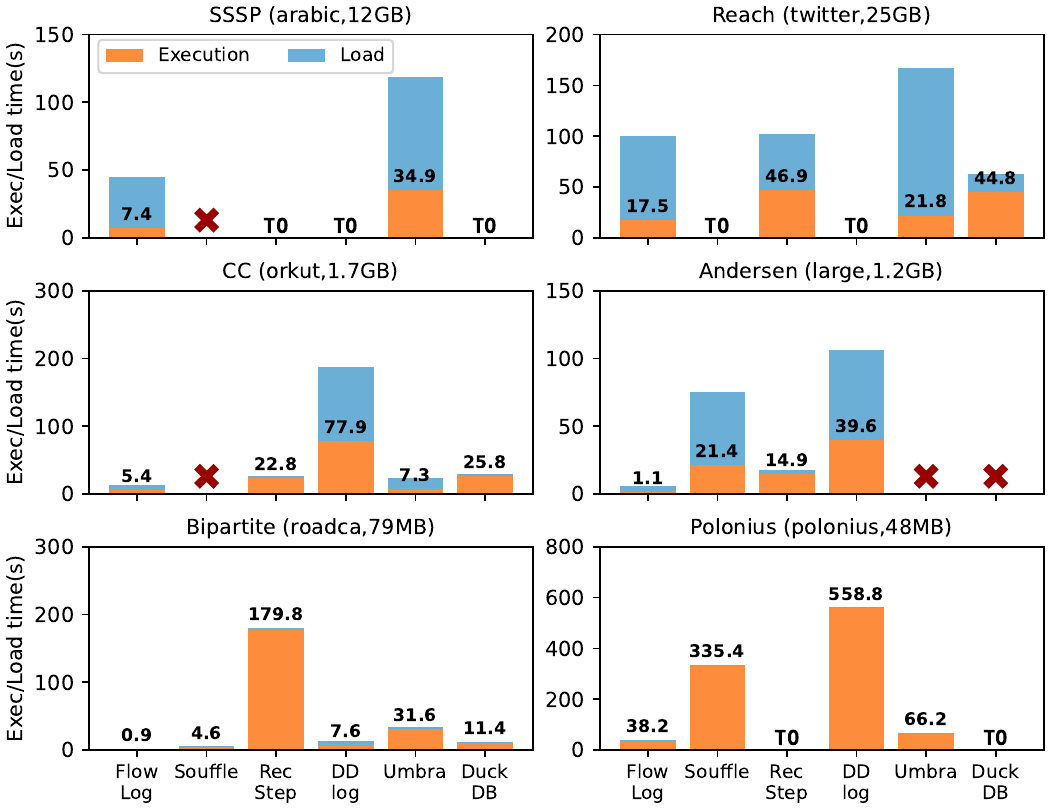}
  \caption{\small 64-thread runtime breakdown (s) for six program-dataset pairs. Stacked bars show data loading (blue) and core \Datalog evaluation (orange); \textcolor{red!60!black}{\Large\textbf{×}} are unsupported cases and \texttt{TO} indicate 900s timeout. Numbers on top of each bar indicate core execution times only.}
  \label{fig:ablation}
\end{figure}

\smallskip
\introparagraph{Ablation Studies} Table~\ref{tab:results} reports end-to-end runtimes; Figure~\ref{fig:ablation} breaks them into CSV loading (blue) and core \Datalog evaluation (orange) for six workloads (large inputs: top two; medium: middle two; small: bottom two). \eclair currently applies no ingestion-specific optimizations for CSV, so its loading time is generally higher than \recstep and DuckDB. In contrast, \souffle and \ddlog exhibit long loading phases because they insert tuples one at a time into single-threaded indexed data structures (e.g., B-trees) before execution; for \textsf{Andersen} (1.2 GB input), their loading takes 54.0s and 66.7s, while \eclair uses 4.6s. Umbra can also be slow due to converting CSV files into its internal columnar format before execution.

\eclair's scaling-up performance is impressive: for all six benchmarks in Figure~\ref{fig:ablation}, its core execution is always the fastest. On \textsf{Reach} (twitter), although the end-to-end time trails DuckDB due to slower loading, its execution (17.5s) is 2.6$\times$ faster than DuckDB's 44.8s. The smaller-but recursion-intensive bottom two workloads (i.e. execution time dominates) further underscore \eclair's core efficiency.

\smallskip
\introparagraph{Performance Analysis} The consistently superior performance of \eclair stems not from one, but from an ensemble of techniques in this paper. Logic fusion and subplan reuse are key mechanisms for controlling memory overhead in workloads generating explosive intermediate results (e.g., \textsf{TC}/\textsf{SG} on \textsf{G40K}), where nearly all competing engines encounter timeouts or \texttt{OOM} (see Table~\ref{tab:results}). Structural query planning and \sip collectively optimize (and stabilize) execution for recursive multiway joins, such as those in \textsf{Galen} and \textsf{DOOP}. Boolean/algebraic specializations (Sec.~\ref{sec:specialization}-\ref{sec:extensibility}) make \eclair competitive in recursive aggregates (in fact, achieving the best performance for both \textsf{CC} and \textsf{SSSP} in Table~\ref{tab:results}). \eclair's design inherits \textsf{DD}'s exceptional asynchrony, allowing it to excel in \textbf{long-tail workloads} (i.e. many lightweight iterations, typical in program analysis). The next sections provide detailed analysis for some of these findings.


\subsection{CPU and Memory Usage} \label{exp:live}
\begin{figure*}[t]
    \centering
    
    \resizebox{\textwidth}{!}{\input{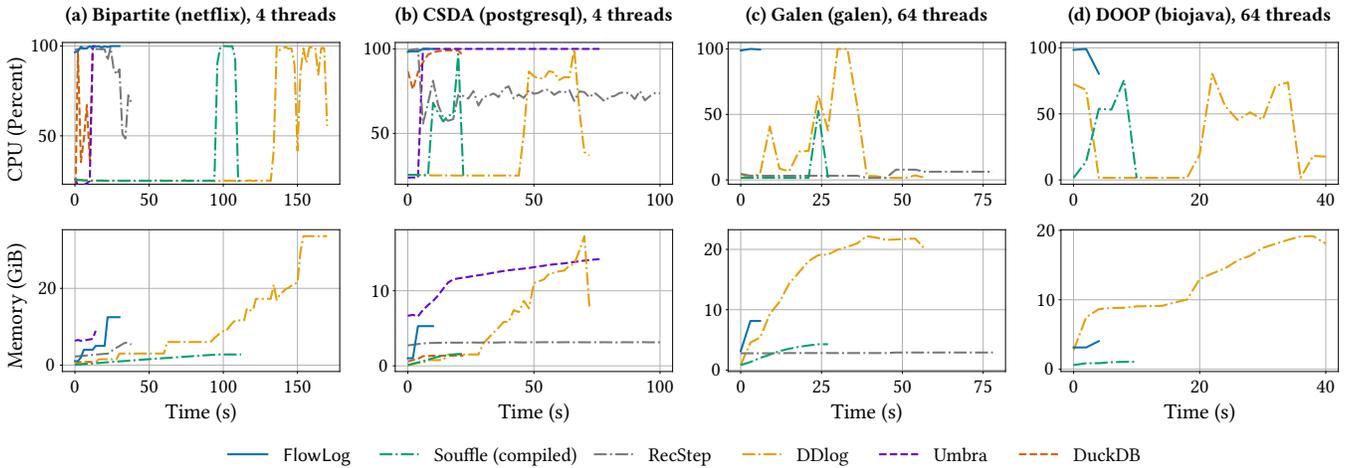}}

    \caption{\small Live CPU (in \%) and memory (in GBs) consumption on 4 different workloads; each panel shows CPU (top) and memory usage (bottom) over its execution horizon. Missing lines indicate unsupported cases (i.e. Umbra/DuckDB on \textsf{Galen}; Umbra/DuckDB/\recstep on \textsf{DOOP}).} 
    \label{fig:live}
\end{figure*}

Figure~\ref{fig:live} reports real-time CPU and memory usage of all systems on four workloads in Table~\ref{tab:results} (\eclair in blue solid curves).

\smallskip

\introparagraph{CPU} For both 4 and 64 threads, \eclair sustains near-100\% CPU. Its high CPU usage stems from \textsf{DD}'s asynchrony, where all computation stages—including \textcolor{blue}{\texttt{iterate}}—remain active and concurrent throughout execution (see Sec.~\ref{sec:plan_execution}), allowing a continuous resource saturation. DuckDB and Umbra, having mature parallelism infrastructures, also generally reach near-peak utilization. In contrast, \souffle employs only outer-loop parallelism in its index nested loop joins; \recstep relies on the DBMS's internal parallelism, but suffers from cross-iteration synchronizations. They both show moderate CPU usage for batch-oriented workloads such as \textsf{Bipartite} (Figure~\ref{fig:live}a, 4 iter.), but degrade sharply—often <25\%—on long-tail workloads such as \textsf{Galen} (Figure~\ref{fig:live}c, 32 iter.). Another inefficiency is evident in Figure~\ref{fig:live}a, where \souffle and \ddlog have long single-threaded phases for initial index construction. On large programs such as \textsf{DOOP}, which feature many rules of hybrid characteristics, \souffle averages <$50\%$ CPU usage. DDlog—lacking dedicated support for parallelism~\cite{DCDatalog22, DBLP:conf/datalog/FanMK22}—displays volatile CPU consumptions.

\smallskip
\introparagraph{Memory} A main inefficiency for \ddlog is its large memory footprint, due to direct usage of \textsf{DD} without careful memory control. In all bottom panels of Figure~\ref{fig:live}, \ddlog exhibits the highest memory usage (>20GB). Umbra also shows elevated usage, especially in long-tail analyses (e.g., Figure~\ref{fig:live}b, 720 iter.) and triggers multiple \texttt{OOM} failures in Table~\ref{tab:results}. This is likely because Umbra does not have DuckDB's \texttt{USING} \texttt{KEY} optimization~\cite{UsingKEY} that inserts new facts in-place, and instead accumulates all results in memory. In contrast, DuckDB, \souffle and \recstep incur lower memory overheads, generally staying <5GB. Although \eclair builds on \textsf{DD}, our optimizations (Sec.~\ref{sec:fusion}–\ref{sec:specialization}) substantially shrink retained states: it uses average 3.5× less memory than \ddlog on these workloads (e.g. 4 v.s. 20 GB on \textsf{DOOP}!), yet still 2–3× more than \souffle's lean baseline.


\subsection{Parallel Scalability}

\begin{figure}[t]
\definecolor{myblue}{RGB}{0,114,178}
\definecolor{mygreen}{RGB}{0,158,115}
\definecolor{mygray}{RGB}{120,120,120}
\definecolor{myorange}{RGB}{230,159,0}
\definecolor{myred}{RGB}{213,94,0}
\definecolor{mypurple}{RGB}{102,0,204}

\begin{tikzpicture}
\begin{groupplot}[
    group style={
        group size=2 by 3,      
        horizontal sep=0.45cm,  
        vertical sep=0.55cm,
    },
    width=0.62\columnwidth,     
    height=4.0cm,
    xmin=1, xmax=64,
    ymin=0,
    xtick={1,2,4,8,16,32,64},
    xmode=log, log basis x=2,
    grid=both,
    grid style={line width=.1pt, draw=gray!20},
    major grid style={line width=.2pt,draw=gray!50},
    tick label style={font=\scriptsize},
    label style={font=\scriptsize, yshift=-0.5em},
    title style={font=\normalsize, yshift=-0.7em},
    every axis plot/.append style={line width=1pt},
    every mark/.append style={solid, scale=0.8, line width=-0.1pt},
]

\nextgroupplot[
    ylabel={\scriptsize Speedup},
    ytick={1,5,10,15,20},
    ymax=21,
]
\addplot+[mygray, mark=diamond*, mark options={fill=mygray}] coordinates {(1,1) (2,1.58) (4,2.83) (8,3.77) (16,5.97) (32,9.72) (64,11.90)};
\addplot+[myorange, mark=*, mark options={fill=myorange}] coordinates {(1,1) (2,1.55) (4,2.23) (8,2.73) (16,3.22) (32,3.49) (64,3.56)};
\addplot+[mypurple, mark=star, mark options={solid}, solid] coordinates {(1,1) (2,1.68) (4,2.55) (8,3.47) (16,4.18) (32,4.80) (64,4.61)};
\addplot+[myred, mark=x, mark options={solid}] coordinates {(1,1) (2,1.53) (4,2.27) (8,3.16) (16,3.67) (32,4.17) (64,3.77)};
\addplot+[myblue, mark=square*, mark options={fill=myblue}, solid] coordinates {(1,1) (2,1.81) (4,3.43) (8,6.10) (16,10.35) (32,17.03) (64,17.92)};
\node[anchor=south west, font=\scriptsize\bfseries, text=black, fill=white, draw=black, rounded corners=3pt, inner xsep=4pt, inner ysep=2pt, align=left]
  at (rel axis cs:0.03,0.6) {\textsf{CC (orkut)}\\
  {14 iterations}\\
  \textcolor{red!60!black}{\Large\textbf{×}} \textsf{Soufflé}};

\nextgroupplot[
    ytick={1,10,20,30,40},
    ymax=41,
    legend to name=group legend,
    legend columns=3,         
    legend style={
      font=\small,
      /tikz/every even column/.append style={column sep=1.2em},
      draw=none,
      at={(0.5,-0.3)},
      anchor=north,
      nodes={inner sep=1pt},
      /tikz/column 2/.style={column sep=1.2em},
    },
]
\addplot+[myblue, mark=square*, mark options={fill=myblue}, solid] coordinates {(1,1) (2,1.89) (4,3.70) (8,7.00) (16,14.02) (32,28.38) (64,36.40)};
\addlegendentry{\footnotesize \eclair}
\addplot+[mygreen, mark=triangle*, mark options={fill=mygreen}] coordinates {(1,1) (2,1.86) (4,3.19) (8,4.80) (16,7.40) (32,10.12) (64,10.95)};
\addlegendentry{\footnotesize Souffle (compiled)}
\addplot+[mygray, mark=diamond*, mark options={fill=mygray}] coordinates {(1,1) (2,0.73) (4,1.06) (8,1.32) (16,1.45) (32,1.84) (64,2.92)};
\addlegendentry{\footnotesize RecStep}
\addplot+[myorange, mark=*, mark options={fill=myorange}] coordinates {(1,1) (2,1.81) (4,3.10) (8,3.67) (16,4.62) (32,5.29) (64,6.77)};
\addlegendentry{\footnotesize DDlog}
\addplot+[mypurple, mark=star, mark options={solid}, solid] coordinates {(1,1) (2,1.84) (4,3.20) (8,5.09) (16,8.19) (32,12.68) (64,12.21)};
\addlegendentry{\footnotesize Umbra}
\addplot+[myred, mark=x, mark options={solid}, solid] coordinates {(1,1) (2,0.99) (4,1) (8,1) (16,1) (32,1) (64,1)};
\addlegendentry{\footnotesize DuckDB}
\node[anchor=south west, font=\scriptsize\bfseries, text=black, fill=white, draw=black, rounded corners=3pt, inner xsep=4pt, inner ysep=2pt, align=left]
  at (rel axis cs:0.03,0.7) {\textsf{TC (G10K-0.001)}\\
  {6 iterations}};

\nextgroupplot[
    ylabel={\scriptsize Speedup},
    ytick={1,5,9,13},
    ymax=14,
]
\addplot+[mygreen, mark=triangle*, mark options={fill=mygreen}] coordinates {(1,1) (2,1.17) (4,1.39) (8,1.52) (16,1.63) (32,1.66) (64,1.69)};
\addplot+[mygray, mark=diamond*, mark options={fill=mygray}] coordinates {(1,1) (2,1.50) (4,2.10) (8,2.80) (16,4.07) (32,4.15) (64,4.53)};
\addplot+[myorange, mark=*, mark options={fill=myorange}] coordinates {(1,1) (2,1.22) (4,1.55) (8,1.63) (16,1.63) (32,1.81) (64,1.84)};
\addplot+[myred, mark=x, mark options={solid}] coordinates {(1,1) (2,1.77) (4,1.96) (8,2.01) (16,2.18) (32,1.87) (64,1.77)};
\addplot+[mypurple, mark=star, mark options={solid}, solid] coordinates {(1,1) (2,1.28) (4,1.50) (8,1.61) (16,1.69) (32,1.74) (64,1.70)};
\addplot+[myblue, mark=square*, mark options={fill=myblue}, solid] coordinates {(1,1.00) (2,1.84) (4,3.41) (8,5.69) (16,8.89) (32,11.65) (64,11.40)};
\node[anchor=south west, font=\scriptsize\bfseries, text=black, fill=white, draw=black, rounded corners=3pt, inner xsep=4pt, inner ysep=2pt, align=left]
  at (rel axis cs:0.03,0.7) {\textsf{Bipartite (netflix)}\\
  {4 iterations}};

\nextgroupplot[
    ytick={1,5,10,15,20},
    ymax=24,
]
\addplot+[mygreen, mark=triangle*, mark options={fill=mygreen}] coordinates {(1,1.00) (2,1.63) (4,2.37) (8,2.75) (16,2.76) (32,3.39) (64,3.34)};
\addplot+[mygray, mark=diamond*, mark options={fill=mygray}] coordinates {(1,1.00) (2,1.11) (4,1.99) (8,2.78) (16,3.89) (32,5.34) (64,6.13)};
\addplot+[myorange, mark=*, mark options={fill=myorange}] coordinates {(1,1.00) (2,1.65) (4,2.14) (8,2.25) (16,2.25) (32,2.57) (64,2.62)};
\addplot+[myblue, mark=square*, mark options={fill=myblue}, solid] coordinates {(1,1.00) (2,1.71) (4,3.21) (8,5.95) (16,10.87) (32,19.57) (64,21.77)};
\node[anchor=south west, font=\scriptsize\bfseries, text=black, fill=white, draw=black, rounded corners=3pt, inner xsep=4pt, inner ysep=2pt, align=left]
  at (rel axis cs:0.03,0.6) {\textsf{CSPA (postgresql)}\\
  {33 iterations}\\
  \textcolor{red!60!black}{\Large\textbf{×}} \textsf{DuckDB/Umbra}};

\nextgroupplot[
    ylabel={\scriptsize Speedup},
    xlabel={\scriptsize Number of Threads},
    ytick={1,4,8,12},
    ymax=15,
]
\addplot+[mygreen, mark=triangle*, mark options={fill=mygreen}] coordinates {(1,1.00) (2,0.94) (4,1.29) (8,1.12) (16,1.09) (32,1.03) (64,1.07)};
\addplot+[mygray, mark=diamond*, mark options={fill=mygray}] coordinates {(1,1.00) (2,1.58) (4,2.64) (8,3.57) (16,4.38) (32,4.86) (64,4.86)};
\addplot+[myorange, mark=*, mark options={fill=myorange}] coordinates {(1,1.00) (2,1.28) (4,1.69) (8,1.77) (16,1.61) (32,1.67) (64,1.90)};
\addplot+[myred, mark=x, mark options={solid}] coordinates {(1,1.00) (2,1.90) (4,3.45) (8,5.33) (16,6.75) (32,7.84) (64,7.19)};
\addplot+[mypurple, mark=star, mark options={solid}, solid] coordinates {(1,1) (2,1.63) (4,3.53) (8,7.05) (16,11.23) (32,13.84) (64,13.30)};
\addplot+[myblue, mark=square*, mark options={fill=myblue}, solid] coordinates {(1,1.00) (2,1.70) (4,3.06) (8,4.97) (16,7.40) (32,8.93) (64,8.24)};
\node[anchor=south west, font=\scriptsize\bfseries, text=black, fill=white, draw=black, rounded corners=3pt, inner xsep=4pt, inner ysep=2pt, align=left]
  at (rel axis cs:0.03,0.7) {\textsf{CSDA (postgresql)}\\
  {720 iterations}};

\nextgroupplot[
    xlabel={\scriptsize Number of Threads},
    ytick={1,5,10,15,20},
    ymax=17,
]
\addplot+[mygreen, mark=triangle*, mark options={fill=mygreen}] coordinates {(1,1.00) (2,1.28) (4,1.38) (8,0.70) (16,0.87) (32,0.79) (64,0.86)};
\addplot+[myorange, mark=*, mark options={fill=myorange}] coordinates {(1,1.00) (2,1.45) (4,2.31) (8,2.45) (16,2.58) (32,2.96) (64,2.87)};
\addplot+[mypurple, mark=star, mark options={solid}, solid] coordinates {(1,1) (2,2.02) (4,4.46) (8,5.84) (16,3.50) (32,4.63) (64,4.71)};
\addplot+[myblue, mark=square*, mark options={fill=myblue}, solid] coordinates {(1,1.00) (2,1.83) (4,3.23) (8,5.70) (16,8.61) (32,13.13) (64,16.14)};
\node[anchor=south west, font=\scriptsize\bfseries, text=black, fill=white, draw=black, rounded corners=3pt, inner xsep=4pt, inner ysep=2pt, align=left]
  at (rel axis cs:0.02,0.6) {\textsf{Polonius (polonius)}\\
  {1487 iterations}\\
  \textcolor{red!60!black}{\Large\textbf{×}} \textsf{RecStep/DuckDB}};

\end{groupplot}
\end{tikzpicture}

\pgfplotslegendfromname{group legend}

\caption{\small Scalability of all competing systems on six program–data pairs; each subplot shows speedups relative to the single-thread run, up to $2^6\!$ threads. Legends give the program–data pair and iterations to converge. Unsupported and timed-out cases are marked by \textcolor{red!60!black}{\Large\textbf{×}}.}
\label{fig:scale}
\end{figure}
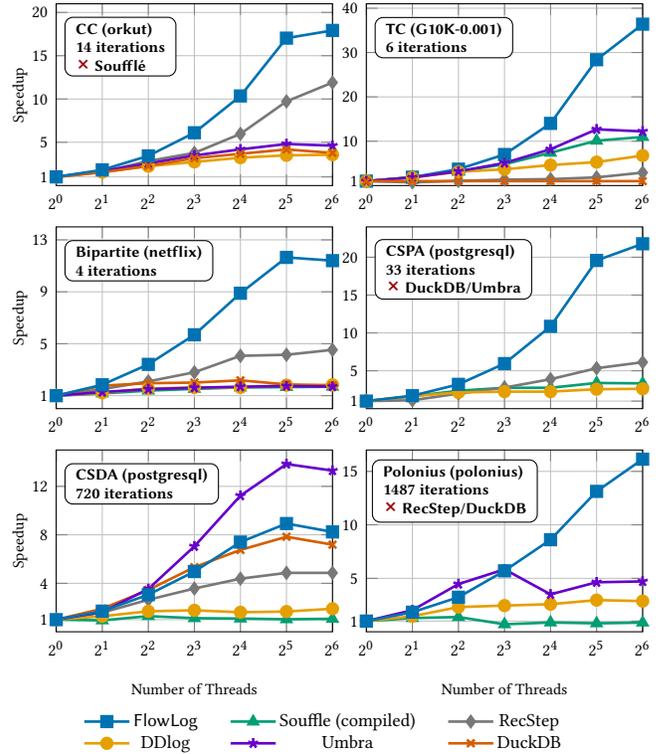

Scalability of \Datalog engines is strongly workload dependent. Figure~\ref{fig:scale} presents the parallel speedups (over single-thread execution) of all systems across six representative cases: three graph queries (top row), and three program analyses (bottom row). \eclair exhibits the most consistent scaling in all workloads: its speedup rises steadily through 32 threads and only mildly tapers at 64. This superior outscaling over the other engines (including DuckDB and Umbra) makes \eclair the most competitive at high thread counts: at 64 threads, \textsf{TC} is the only one in six where another engine (i.e. Umbra, 6.1s) narrowly outperforms \eclair (8.5s, a 38× speedup over its own baseline); DuckDB has almost no scaling in this case.

On long-tail workloads, parallelism may degrade; and overheads of thread management and data exchange may outweigh the benefits. We see this in \textsf{CSDA} (postgresql, 720 iterations) and \textsf{Polonius} (1487 iterations): \souffle and \ddlog gain almost no parallel speedup. In \textsf{CSDA}, DuckDB and \recstep scale slightly weaker and Umbra outscales \eclair, but they all start from a much slower single-thread baseline than \eclair. This is largely due to these databases' lack of continuously maintained views, forcing them to re-compute them at every iteration~\cite{DBLP:conf/datalog/FanMK22}. Umbra's occasionally competitive scaling (e.g. \textsf{CSDA}) is due to its morsel-driven parallelism~\cite{LeisBK014}, which allows it to adaptively balance work across threads.


\subsection{Structural Planning and Robust Execution}
\label{sec:planvar}

\begin{figure*}[!htbp]
  \centering

  \begin{minipage}[t]{0.65\textwidth}
    \vspace{0.7pt}
    \resizebox{\textwidth}{!}{\input{figures/plan_variance.pgf}}
  \end{minipage}
  \hfill
  \begin{minipage}[t]{0.34\textwidth}
    \vspace{0pt}
    \footnotesize  
    \setlength{\tabcolsep}{4pt}\renewcommand{\arraystretch}{1.15}
    \begin{tabular}{@{}l|rrrc@{}}
      \toprule
      Prog, rule & plan+\texttt{sip}\textsuperscript{$\star$} & plan only & \texttt{sip} only & no opt. \\
      \midrule
      \multirow{3}{*}{z3, $r_{18}$}
        & 22.6s & 19.3s & 22.2s & 21.5s  \\
        & 20.5s & 19.7s & 22.0s & 20.5s \\
        & 19.9s & 20.8s & {\scriptsize \faSkullCrossbones}\texttt{OOM} & {\scriptsize \faSkullCrossbones}\texttt{OOM}  \\ \hline
      \multirow{3}{*}{batik, $r_{118}$}
        & 14.6s & 12.4s & 15.1s & 11.7s  \\
            & 21.1s & {\scriptsize \faSkullCrossbones}343.6s & 20.5s & {\scriptsize \faSkullCrossbones}346.5s  \\
        & 21.4s & {\scriptsize \faSkullCrossbones}343.7s & 18.7s & 16.7s  \\ \hline
      \multirow{3}{*}{galen, $r_{2}$}
        & 8.8s & 7.4s & 8.5s & 7.4s  \\
        & 8.0s & 6.8s & {\scriptsize \faSkullCrossbones}\texttt{OOM} & {\scriptsize \faSkullCrossbones}\texttt{OOM}  \\
        & 10.2s & 11.2s & {\scriptsize \faSkullCrossbones}\texttt{OOM} & {\scriptsize \faSkullCrossbones}\texttt{OOM}  \\
      \bottomrule
    \end{tabular}
  \end{minipage}%

\vspace{0pt}
\noindent\begin{tabular*}{\textwidth}{@{}p{0.63\textwidth}@{\hspace{1.5em}}p{0.34\textwidth}@{}}
{%
  \captionsetup{font=small,aboveskip=0pt,belowskip=0pt}
  \definecolor{darkgreen}{rgb}{0.0, 0.5, 0.0}
  \captionof{figure}{Runtime variability (64 threads) taking different listing orders.
  \textcolor{darkgreen}{Green} numbers and dotted lines show the medians for \eclair using the join optimizer and \texttt{sip}, \eclair disabling both optimizations, and \souffle.
  \texttt{TO} (i.e. >900s) and \texttt{OOM} cases are marked ×.}%
  \label{fig:planvar} 
}
&
{%
  \captionsetup{font=footnotesize,aboveskip=0pt,belowskip=0pt}
  \captionof{table}{Runtime variability (64 threads) taking every rule ordering for $r_{18}$ in \textsf{DDISASM}, $r_{118}$ in \textsf{DOOP} (batik), and $r_{5}$ in \textsf{Galen}, comparing \eclair (plan+\texttt{sip}), and \eclair disabling the query planner, \texttt{sip}, or both.}%
  \label{tab:program-times} 
}
\end{tabular*}

\end{figure*}

Next, we will study how join ordering impacts \Datalog performance, and empirically validate our two complementary techniques: robust execution via semijoin prefiltering (or \sip for short, see Sec.~\ref{sec:robust}) and structural (worst-case) query planning (see Sec.~\ref{sec:planning}).

First, we revisit \recstep's reliance on its \dbms optimizer to optimize join orders on the fly. In programs such as \textsf{Reach} and \textsf{Andersen}, where the optimal plan is either obvious (e.g., a single join) or plan choices have little performance variance, \recstep repeatedly invokes the optimizer at each iteration only to regenerate the same plan. This overhead erodes \recstep's overall performance (Table~\ref{tab:results}). \textsf{Galen} further exposes its shortcomings: as data skew shifts across iterations, \recstep does not pivot promptly to a better order, spends many iterations in slow plans, causing eventual time-out.

Now, we demonstrate that \eclair's techniques in Sec.~\ref{sec:planning} and \ref{sec:robust} jointly make it performant and robust. Figure~\ref{fig:planvar} and Table~\ref{tab:program-times} report runtime and plan sensitivity for three benchmarks having recursive multi-way joins: \textsf{Galen}, \textsf{DOOP} (batik), and \textsf{DDISASM} (z3). 

\smallskip
\introparagraph{Performance and Robustness (Figure~\ref{fig:planvar})} For each benchmark, we repeatedly sample rules (having recursive multiway joins) and rewrite their listing orders into new, unused variants. We exclude variants that introduce cross products, since \souffle executes them verbatim and typically triggers time-outs. Figure~\ref{fig:planvar} compares runtime across: $(i)$ default \eclair (having techniques from Sec.~\ref{sec:planning}-\ref{sec:robust}), $(ii)$ \eclair turning off both techniques, and $(iii)$ \souffle. To distinguish, we will refer to them as \eclair (plan+\sip), \eclair (no opt.) and \souffle for the rest of this section. We omit \ddlog, as \eclair (no opt.) can be regarded as a memory-optimized variant of it. DuckDB and Umbra lack support for these programs.

Figure~\ref{fig:planvar} shows that both \eclair (no opt.) and \souffle are highly sensitive to join ordering. Across 91 distinct listing orders, \eclair (no opt.) times out or \texttt{OOM} on 25 instances, and \souffle times out on 23—over 25\% of all cases! In \eclair (no opt.), bad orders causes \textsf{DD} to maintain large incremental join results, leading to timeouts/\texttt{OOM}. \souffle, while avoids materialization via pipelined execution (see Sec.~\ref{sec:plan_execution}), still suffers from long runtimes due to expensive nested-loop joins. In contrast, \eclair (plan+\sip) never times out or runs \texttt{OOM} on any plan, and its runtime distributions are much more stable. 

The figure also reports median runtimes over all evaluated join orders (\texttt{TO}/\texttt{OOM} conservatively set to 900s, favoring \eclair (no opt.) and \souffle). In every benchmark, \eclair (plan+\sip) achieves the lowerest median runtime. The largest gains appear in \textsf{DOOP}: 2.4× over \eclair (no opt.) and 9.5× over \souffle. These speedups stem from expensive multi-way joins where the structural planner (Sec.~\ref{sec:planning}) selects substantially better—though not always optimal—plans than fixed listing orders. For example, for the rule of Example~\ref{ex:doop}, poorly assigned orders cause both \eclair (no opt.) and \souffle to time out, whereas \eclair (plan+\sip)'s optimizer chooses the bushy plan in Figure~\ref{fig:doop} (right), completing the entire program in 11s.


\smallskip
\introparagraph{Planning \& \sip are complementary (Table~\ref{tab:program-times})} Structural planning often steers away from pathological join orders, but for complex multi-way joins where the optimal order remains elusive, \sip prefilters dangling tuples and cushions poor structural choices. For example, taking the bad listing order of $r_3$ in Example~\ref{ex:galen}, \eclair (no opt.) takes 186s to run and peaks at 144 GB; \eclair (plan+\sip), even though the structural optimizer cannot find a better order, uses \sip pruning to finish in 86s with a 47~GB peak.


Table~\ref{tab:program-times} further dissects this finding, showing that both planning and \sip are necessary for \eclair to be fast and robust. For each benchmark in Figure~\ref{fig:planvar}, we pick three representative recursive rules ($r_{18}$, $r_{118}$, $r_{2}$) whose bodies are triangular (three-way) recursive joins; for each such rule we enumerate all three distinct binary join orders. The table reports \eclair's runtimes under four settings: full system, planner only, \sip only, and both optimizations disabled. Skull face entries highlight massive slowdowns or memory blow-ups. We observe that: $(i)$ the structural planner alone sometimes lands on pathological orders that inflates intermediates; $(ii)$ \sip alone adds little benefit on non-selective joins and makes no attempt to improve the join order (note that it also incurs moderate semijoin overheads, albeit by a small margin); and $(iii)$ using both lets \sip prefilter inputs so the remaining joins behave close to a worst-case data distribution (well handled by the worst-case-oriented planner), yielding the overall fastest and most robust column in Table~\ref{tab:program-times}.

\section{Conclusion}

We developed \eclair, a new Datalog engine that decouples the logical optimizations from the physical execution on DD's operators. \eclair’s IR-level planning and semijoin prefiltering cut memory, stabilize recursion, and deliver up to order-of-magnitude speedups over \ddlog, \souffle, \recstep, \ddlog, DuckDB, and Umbra across diverse benchmarks. Looking ahead, we plan to add robust cardinality estimation, cost-aware optimizers, compile-time optimizations, richer incremental \Datalog features, and elastic scale-out execution.


\begin{acks}
 We gratefully acknowledge Sam Arch, Frank McSherry, Kristopher Micinski, Thomas Neumann, and Yihao Sun for their stimulating discussions and helpful insights, which greatly shaped and inspired the design of \eclair.
\end{acks}


\bibliographystyle{ACM-Reference-Format}
\bibliography{refs}

\end{document}